# Quantum compressed sensing

Jianyong Hu[1,4,7*†], Wei Li[1,4†], Shuxiao Wu[1,4*], Liwen Zhang[2], Yongchuang Sun[1,4], Jiazhao Tian[3], Guosheng Feng[5], Zhixing Qiao[5], Jianqiang Liu[6], Changgang Yang[1,4], Ruiyun Chen[1,4], Chengbing Qin[1,4], Guofeng Zhang[1,4], Liantuan Xiao[1,3,4,7*] and Suotang Jia[1,4]

*[1]State Key Laboratory of Quantum Optics Technologies and Devices, Institute of Laser Spectroscopy, Shanxi University, Taiyuan 030006, China*

*[2]School of Physics and Information Engineering, Shanxi Normal University, Taiyuan 030031, China*

*[3]College of Physics and Optoelectronics Engineering, Taiyuan University of Technology, Taiyuan 030024, China*

*[4]Collaborative Innovation Center of Extreme Optics, Shanxi University, Taiyuan 030006, China*

*[5]College of Medical Imaging, Shanxi Medical University, Taiyuan, 030001, China*

*[6]College of Information Engineering, Shanxi Vocational University of Engineering Science and Technology, Jinzhong 030619, China*

*[7]Hefei National Laboratory, Hefei 230088, China*

**Corresponding author E-mail address: jyhu@sxu.edu.cn; wushuxiao1@sxu.edu.cn; xlt@sxu.edu.cn*

†*These authors contributed equally to this work.*

**Abstract:** How many measurements are fundamentally required to capture a signal? Shannon's information theory established the bedrock of this question in 1948, the Nyquist–Shannon theorem set the first answer, and compressed sensing (CS) rewrote it in 2006 by reducing the required measurement number to $M = O(K\log(N/K))$ for a $K$-sparse signal. Here, we propose quantum compressed sensing (QCS), a paradigm that reframes signal acquisition as a unitary quantum evolution. By encoding high-dimensional signal information into a single quantum probe state, then introducing domain-alignment evolution—a physically realizable unitary transformation that maps the sparse basis directly onto the measurement basis. QCS executes the support-set search at the quantum level without consuming measurement trials. The logarithmic penalty vanishes, compressing the required measurement number from the classical bound to $M \sim K$ and reducing reconstruction from ill-posed optimization to linear estimation. We experimentally validate QCS using frequency- and time-domain sparse signals, confirming that the measurement number scales linearly with sparsity and decouples entirely from the signal dimension. Our work provides a physical pathway toward ultimate information acquisition efficiency, with broad implications for sensing, imaging, and communication.

## Introduction

The development of signal sampling theory is essentially a history of constantly questioning how many measurements are needed at least to faithfully restore the signal. The Nyquist-Shannon theorem requires sampling at twice the bandwidth of signal for reliable reconstruction (*1, 2*). Compressed sensing (CS) offered a breakthrough by showing that if a signal is sparse in some transform domain, it can be recovered from far fewer non-adaptive linear projections than the Nyquist rate (*3*). This shifted the sampling efficiency constraint from bandwidth to signal structure (*4*).

Nevertheless, classical CS encounters a theoretical boundary in its sampling efficiency. From an information-theoretic perspective, the entirety of information in a $K$-sparse signal is completely determined by merely $K$ parameters (*5, 6*). With prior knowledge of the signal's support set, one can reconstruct the signal using only $K$ measurements, which represents the ideal sampling efficiency promised by adaptive sampling strategies (*7-10*). However, the non-adaptive measurement of classical CS precludes any prior knowledge of the support set before measurements are taken. The measurement matrix treats all possible signal uniformly to satisfy the universality required by the restricted isometry property (RIP) (*11*). Blindly searching for $K$ nonzero components within an $N$-dimensional space incurs a logarithmic combinatorial search cost of $\log(N/K)$, which sets the lower bound on the measurement number at $M = O(K\log(N/K))$ in classical CS (*3*, *11*, *12*) (see Supplementary Material S1).

Quantum technology provides a pathway to breach this boundary (*13-16*). The inherent parallelism of quantum superposition states enables a single quantum state to simultaneously explore all possible signal space, thereby encoding the complete information of a high-dimensional signal. Correspondingly, the collapse outcome of quantum projective measurements is governed by the probability distribution established by the system's evolution. This suggests that the combinatorial search cost of classical CS can be circumvented by leveraging quantum parallelism and the statistical nature of projective collapse (*17-20*).

Here, we propose a quantum compressed sensing (QCS) paradigm that transforms the mathematical reconstruction problem into a quantum physical evolution process. This paradigm compresses the required measurement number from the logarithmic scaling of $O(K\log(N/K))$ to a constant-order scaling of $M \sim K$, approaching the theoretical bound of adaptive sampling. The conceptual core of QCS is to reframe signal acquisition as a quantum physical evolution. By exploiting the parallelism of quantum superposition, a single probe state can carry the complete information of a high-dimensional signal. A domain-alignment evolution then unitarily maps the sparse basis directly onto the measurement basis. This operation physically executes the support-set search without consuming any measurement trials, so that subsequent projective measurements naturally concentrate on the nonzero sparse coefficients. Reconstruction thereby reduces from an ill-posed combinatorial optimization to a linear statistical estimation, with the computational complexity dropping from polynomial scaling to $O(K)$. Experimentally, we validate this paradigm using both frequency-domain and time-domain sparse signals, confirming that it compresses the measurement number from the classical lower bound to $M \sim K$, demonstrating a quantum advantage over classical CS. By offloading the combinatorial search from algorithms to unitary evolution, QCS establishes a foundational framework for information acquisition at its physical limit, with generality that extends across classical and quantum sensing disciplines.

**General quantum compressive sensing paradigm**

QCS is defined as a method that encodes high-dimensional signal information into a quantum superposition state, maps the sparse basis to the measurement basis through physical evolution, and focuses projective measurements onto the nonzero sparse coefficients, thereby reconstructing transform-domain sparse signals with sampling efficiency approaching the adaptive sampling limit. It is important to note that the QCS framework presented here is conceptually distinct from prior works that employ classical CS techniques for quantum state tomography (*21-23*).

The signal $x \in \mathbb{R}^N$ of interest is $K$-sparse and can be expressed as $x = \Psi S$, where $\Psi$ is the sparse basis and $S$ is the sparse coefficient vector (*24*). The QCS paradigm comprises four steps.

***Step 1: Quantum probe state preparation.*** A quantum probe state $|\psi_0\rangle$ is prepared to serve as the carrier for high-dimensional signal encoding. The probe state must satisfy two basic requirements. First, its native measurement basis must match the original domain of the signal, ensuring direct linear mapping. Second, it must also exhibit a uniform probability distribution across this basis to guarantee universal sensing capability for arbitrary input signals. The choice of the probe state depends on the native domain of the signal (*25, 26*). For time-domain radio-frequency signals, a coherent state $|\alpha\rangle$ provides a natural choice. Its native measurement outcomes are resolved in the time domain, which matches the original domain of the signal. Moreover, the coherent state possesses intrinsic randomness in the time basis, rendering it incoherent with the Fourier basis and thereby satisfying the restricted isometry property with high probability, see Supplementary Material S2.3. For spatial image signals, spatially structured coherent states or entangled states could be employed.

***Step 2: Linear signal-to-quantum-state mapping.*** In this step, the signal is loaded onto the quantum probe state such that the modulus squared of the wavefunction evolves in direct proportion to the signal amplitude. This linearity ensures that the entire QCS process can be represented as a linear system, a prerequisite for the subsequent quantum collapse to directly reflect the sparse coefficients. Specifically, a signal-controlled evolution operator $\hat{U}_x$ is constructed such that the initial probe state $|\psi_0\rangle$ is mapped to $|\psi_x\rangle=\hat{U}_x|\psi_0\rangle$, with the output state depending linearly on the signal amplitude $x$. Depending on the physical platform, such linear mapping can be realized through intensity modulation, phase encoding, or spatial modulation, see Supplementary Material S2.1. In the present experiments, we employ a broadband electro-optic intensity modulator to linearly load a radio-frequency signal onto a coherent optical field.

***Step 3: Domain-alignment evolution.*** In this step, a physically realizable unitary operator $\hat{U}_e$ is introduced to map the sparse basis onto the measurement basis, thereby establishing a direct correspondence between the signal's sparse domain and the detector's measurement domain. Under this mapping, the collapse outcomes of projective measurements would directly reflect the sparse structure of the signal. The process transforms the sparse localization problem that classical CS addresses algorithmically into a physical evolution of the quantum system.

Let the sparse basis be $\{|\psi_n\rangle\}$ and the detector's measurement basis be $\{|\omega_n\rangle\}$. When these two bases are naturally identical, no additional operation is required. When they differ, the evolution operator $\hat{U}_e$ is constructed to satisfy

$$\left|\psi_{out}\right\rangle = \hat{U}_e\left|\psi_x\right\rangle = \sum_{n=1}^{N} s_n\left|\omega_n\right\rangle, \quad \hat{U}_e = \sum_{n=1}^{N}\left|\omega_n\right\rangle\left\langle\psi_n\right|. \tag{1}$$

This condition establishes a unitary transformation from the sparse domain to the measurement domain at the physical level. Under this architecture, the probability $P(n)$ of detecting a photon in measurement basis $|\omega_n\rangle$ becomes directly related to the corresponding signal's sparse coefficient $s_n$. In an experiment, $\hat{U}_e$ may be realized by any physical process capable of implementing the required basis transformation, such as a programmable quantum gate sequence, a diffractive neural network, or a dispersive optical system (*27*). After this evolution, the output state $|\psi_{out}\rangle$ carries a probability distribution that directly corresponds to the sparse coefficients, and it proceeds to the detection and reconstruction step.

***Step 4: Quantum state detection and signal reconstruction.*** In the quantum state detection substage, the projective measurements are performed on the output state $|\psi_{out}\rangle$ in the measurement basis $\{|\omega_n\rangle\}$. According to the Born rule, the probability of finding the system in state $|\omega_n\rangle$ is

$$P(n) = \left|\left\langle\omega_n\middle|\psi_{out}\right\rangle\right|^2 = \left|\left\langle\omega_n\middle|\hat{U}_e\hat{U}_x\middle|\psi_0\right\rangle\right|^2. \tag{2}$$

After domain-alignment evolution, $|\psi_{out}\rangle$ carries a probability distribution that directly corresponds to the sparse coefficients, so each projective measurement is equivalent to one adaptive sample of a sparse component. By repeating the prepare-evolve-measure cycle $M$ times, we obtain a measurement vector $y = \{y_1, y_2, ..., y_m, ..., y_M\}$.

In the signal reconstruction substage, the measurement vector $y$ is used to estimate the original signal $x$. Under ideal domain alignment, the detection probability $P(n)$ and the target sparse coefficient $s_n$ satisfy a linear relationship $P(n) = \eta s_n$, where $\eta$ is a global scaling factor that can be determined experimentally. By counting the number of detections $k_n$ at measurement basis state $|\omega_n\rangle$ over $M$ detection events, we obtain the estimated probability

$$\hat{P}(n) = \frac{k_n}{M} \tag{3}$$

The sparse coefficient estimates follow directly as $\hat{s}_n = \hat{P}(n)/\eta$. The original signal is then reconstructed via the linear inverse transform $\hat{x} = \Psi\hat{S}$.

The reconstruction paradigm of QCS differs fundamentally from that of classical CS. Classical CS can be viewed as solving a jigsaw puzzle. Each measurement provides a partial mixture of signal components, and an algorithm must invert an ill-posed problem under a sparsity constraint (*28*). QCS, in contrast, operates as a voting process. Domain alignment evolution maps the sparse basis onto the measurement basis, so projective measurements naturally concentrate on the nonzero sparse coefficients. Reconstruction thereby reduces to statistical counting of measurement outcomes. Physical evolution replaces numerical search, converting a high dimensional inverse problem into linear parameter estimation (*29*). This shift yields two fundamental advantages.

First, the measurement number $M$ is no longer constrained by the logarithmic cost of combinatorial search. It depends only on the sparsity $K$ and the desired estimation accuracy, enabling sampling efficiency that approaches the adaptive limit of $M \sim K$. Second, the computational complexity of reconstruction drops from the $O(KMN + K^3)$ scaling typical of classical methods to $O(K)$. A detailed comparison of CS and QCS is provided in Supplementary Materials S2.2.

The four-step architecture of QCS is not a fixed, unalterable sequence but a modular and configurable methodology. Steps 1 and 2 together form a quantum front-end that serves as a universal, incoherent sampling engine whose measurement matrix inherits the restricted isometry property from the intrinsic randomness of the quantum probe. Step 3, domain alignment evolution, could be a decoupled module. Beyond physical domain alignment, it can also be substituted by classical post-processing, which retains the sampling advantages conferred by the quantum front-end while offering an immediate, high-fidelity reconstruction pathway. Step 4, for tasks that do not require complete signal reconstruction, such as signal classification or feature identification, decisions can be made directly from the estimated sparse vector $\hat{S}$ without the inverse transform.

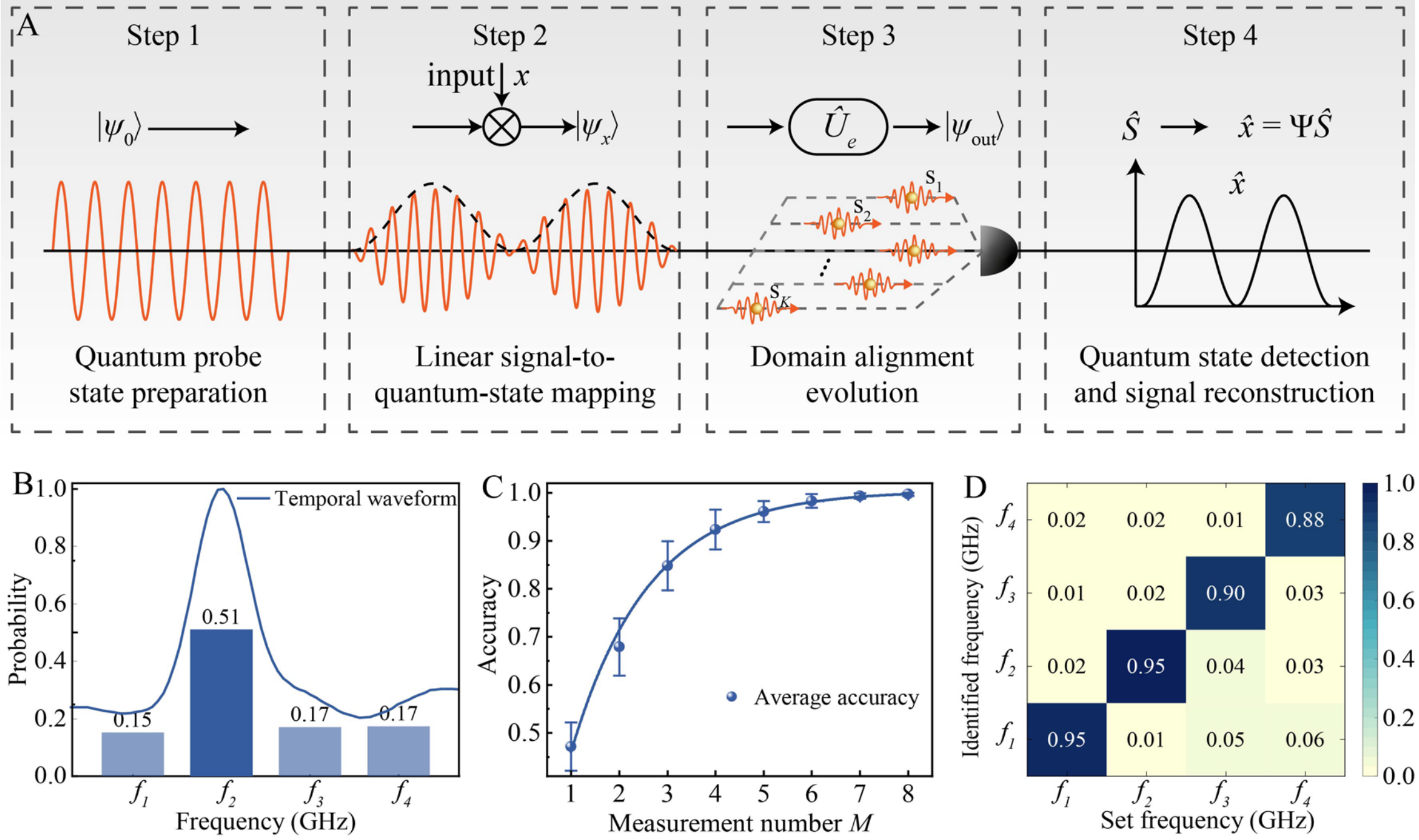

**Figure 1. QCS paradigm and frequency-domain validation**. (A) The QCS workflow comprises four steps. Step 1, quantum probe state preparation. Step 2, linear signal-to-quantum-state mapping. Step 3, domain-alignment evolution. Step 4, quantum state detection and signal reconstruction. (B–D) Frequency-domain sparse signal reconstruction using time-lens-based domain alignment. (B) Probability distribution over the sparse vector for a single-tone sinusoidal signal ($K = 1$, frequency 16.2 GHz) after domain alignment evolution, where $f_1$=5.4 GHz, $f_2$=16.2 GHz, $f_3$=27.0 GHz, $f_4$=37.8 GHz. (C) Sparse component identification accuracy versus measurement number $M$ (i.e., photon counts). (D) Confusion matrix based on four-photon detection events. Horizontal axis denotes the true signal frequency, vertical axis denotes the identified frequency.

From a quantum information theoretic perspective, the advantage of QCS originates from a fundamental shift in where the combinatorial search for the support set is performed. In classical CS, the logarithmic cost $\log(N/K)$ is paid by the measurement data themselves because each non-adaptive measurement acquires only local mixed information. QCS offloads this cost to the quantum evolution stage. A single superposition state traverses all sparse configurations in parallel before measurement. Domain alignment evolution then unitarily maps the sparse basis onto the measurement basis, physically executing the support set search without consuming any measurement trials. When projective measurement finally occurs, the pre-aligned probability distribution steers the collapse toward the nonzero components. Each detection event thus constitutes an adaptive sample that requires no prior knowledge, and the measurement number decouples from the signal dimension $N$, compressing from $M = O(K \log(N/K))$ to $M \sim K$. A detailed quantum information-theoretic analysis is provided in Supplementary Material S2.4.

**Results and discussion**

We validated the QCS paradigm using two complementary experimental configurations. In the first, the signal's sparse domain and the detector's measurement domain differ, requiring active construction of the domain alignment operator. In the second, the two domains naturally coincide, allowing the adaptive sampling mechanism to be characterized without additional physical evolution. Together, these configurations test the universality of the QCS framework and quantitatively benchmark its sampling efficiency.

***Frequency-domain sparse signal measurement***

A defining feature of the QCS paradigm is that its steps can be flexibly configured for different application demands. The frequency-domain sparse signal measurement consists of two complementary paths, which jointly verify the effectiveness of the QCS paradigm under different implementation strategies. One is the domain-aligned direct recognition path based on the time-lens spectrometer (TLS), which fully executes the four-step QCS process and physically confirms the theoretical limit of sampling efficiency. The other is the post-processing reconstruction path based on the Discrete Fourier Transform (DFT), which shifts the domain alignment step from physical evolution to classical post-processing, highlighting an important feature of the QCS paradigm, that each step can be independently tailored to meet engineering constraints. In this path, the incoherent sampling data based on quantum intrinsic randomness generated by steps 1 and 2, even after classical spectral estimation processing, still shows significantly better reconstruction quality than traditional uniform sampling, which indirectly confirms the independent contribution of the QCS's preceding steps.

***TLS domain alignment scheme*;** The signal under test $x(t) \in \mathbb{R}^N$ ($t \in [0, T]$) is a $K$-sparse signal in the frequency domain. The experimental principle is shown in Fig. 1(A). Following the four-step QCS protocol, a coherent state $|\alpha\rangle$ was prepared as the quantum probe state and the radio-frequency signal $x(t)$ was linearly loaded onto the optical field via an electro-optic intensity modulator. Domain alignment was achieved using a TLS, which physically implements a time-to-frequency mapping (*30*). The TLS comprises an electro-optic phase modulator and a dispersion-compensating fiber. The phase modulator is driven by a periodic quadratic phase signal $\varphi(t) \propto t^2$. When the imaging condition between the time-lens chirp rate and the fiber dispersion is satisfied, the photon detection probability $P(t_n)$ of the output state $|\psi_{out}\rangle$ in the time domain became linearly related to the power spectrum $|S(f_n)|^2$ of the input signal within the corresponding time window $T$

$$P(t_n) = \left|\langle t_n | \psi_{out} \rangle\right|^2 = \left|S(f_n)\right|^2, \tag{4}$$

with the time-to-frequency mapping given by $f_n = -t_n / (2\pi\ddot{\phi})$, where $\ddot{\phi}$ denotes the second-order dispersion of the dispersion-compensating fiber.

The output state $|\psi_{\text{out}}\rangle$ was subjected to $M$ projective measurements, yielding a measurement vector of photon arrival times $y = \{t_1, t_2, ..., t_M\}$. By counting the number of timestamps $k_n$ falling into different time bins (corresponding to different signal frequencies $f_n$), we obtained an estimate of the detection probability $\hat{P}(t_n)=k_n/M$. Since $\hat{P}(t_n)$ is linearly related to the frequency-domain sparse coefficient $s_n = |S(f_n)|^2$, the sparse coefficient estimate follows directly as

$$\hat{s}_n = \frac{\hat{P}(t_n)}{\eta}. \tag{5}$$

Figure 1(B) shows the probability distribution over a four-dimensional sparse vector for a single-tone signal ($K = 1$). The measurement probability at the true frequency is significantly higher than at other frequencies. A full estimate of the sparse vector $\hat{S}$ is thus obtained, and the original signal can be reconstructed via the inverse transform $\hat{x}(t) = \Psi\hat{S}$.

A key advantage of this scheme is that a single detection event already provides frequency information. Figure 1(C) shows that the frequency identification accuracy reaches 47% with a single photon detection event and rises to 92% with four photon detection events, directly reflecting that QCS approaches the adaptive sampling limit. Figure 1(D) presents the confusion matrix for four-photon detection events. The TLS-based domain-alignment scheme transforms the complex iterative reconstruction of classical CS into statistical counting of photon arrival events. The measurement number $M$ depends only on the sparsity $K$ and the desired estimation accuracy, and the computational complexity is reduced to the $O(K)$ level. Detailed experimental parameters are provided in Supplementary Materials S3.1.

***DFT-based scheme*;** To demonstrate the flexibility of the QCS framework and the independent contribution of its preceding steps, we further explored a second implementation in which domain alignment is replaced by classical post-processing. In this approach, the linearly mapped quantum state $|\alpha_x\rangle$ is directly detected, yielding a sequence of photon arrival times $y = \{t_1, t_2, \ldots t_m, \ldots, t_M\}$. Applying a DFT to this sequence yields an estimate of the frequency-domain coefficients of the signal (*31, 32*).

$$\hat{s}_n = \sum_{m=1}^{M} \exp(2\pi f_n t_m),\ m=[1, 2, \ldots, M], \tag{6}$$

where $f_n$ represents the frequency corresponding to the frequency-domain coefficient $s_n$ ($n$=1, 2, …, $N$). The largest $K$ frequency-domain coefficients $s_n$ are selected as the nonzero sparse coefficients $\{\hat{s}_k\}$, and the time-domain signal $\hat{x}(t)$ is reconstructed via an inverse Fourier Transform. Figure 2(A-C) show the reconstruction of a single-tone signal ($K = 1$, frequency 20.0 GHz). Figure 2(A) presents the original signal, Figure 2(B) the reconstructed sparse coefficients, and Figure 2(C) the reconstructed signal. Figures 2(D-F) show the reconstruction of a frequency-domain sparse signal with $K = 830$ and a Gaussian pulse envelope (pulse width 120 ps, repetition rate 10 MHz, bandwidth 8.30 GHz). Figure 2(D) presents the original input signal, Figure 2(E) the reconstructed sparse coefficients, and Figure 2(F) the reconstructed Gaussian pulse. These results indicate that the DFT pathway maintains high fidelity even in complex signal scenarios.

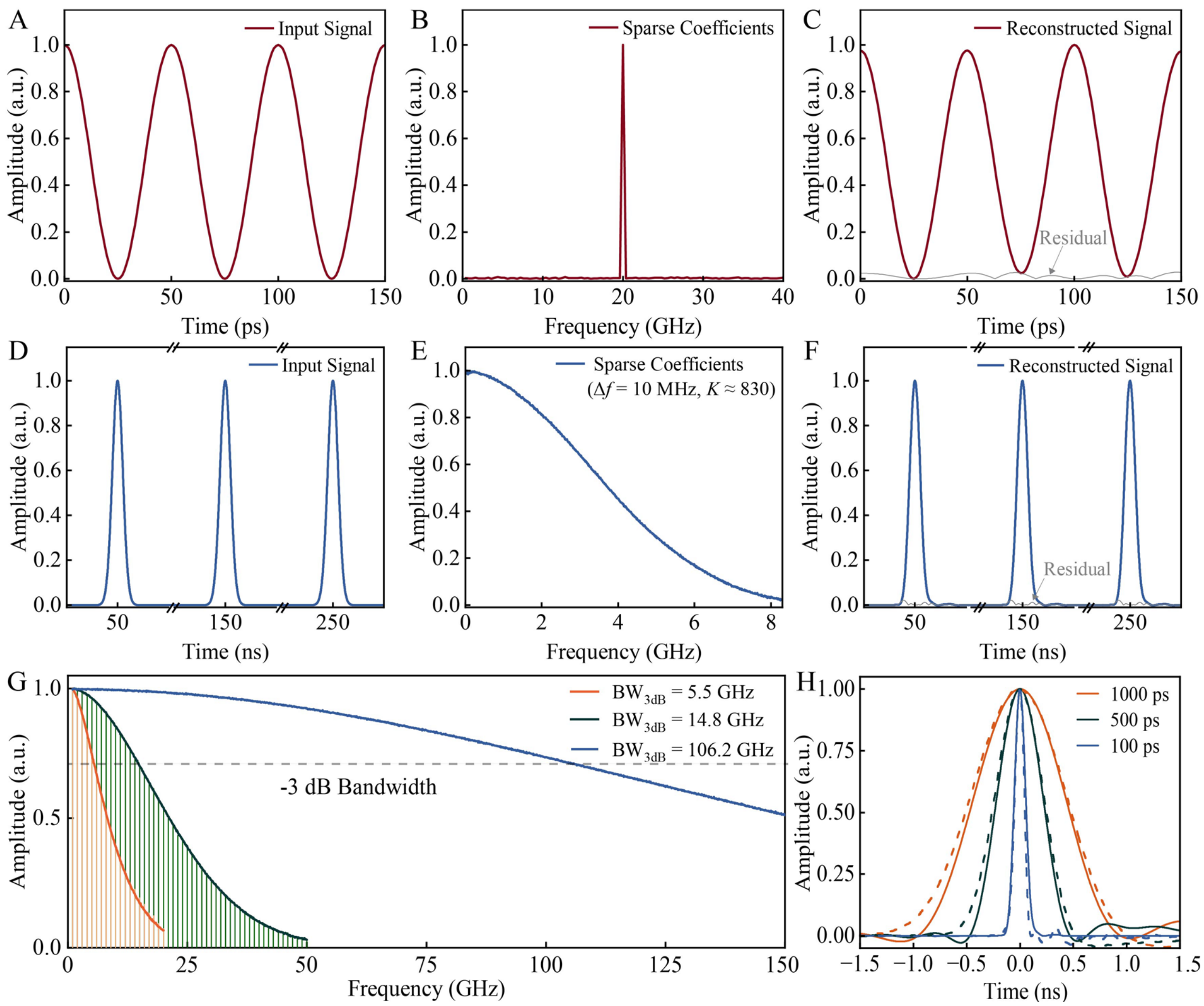

**Figure 2. Frequency-domain sparse signal reconstruction based on DFT scheme**. (A-C) Reconstruction of a single-tone signal ($K$ = 1, frequency = 20.0 GHz). (A) Original input signal $x$. (B) Recovered sparse coefficients $s_n$. (C) Reconstructed signal $\hat{x}$. The residual error between the reconstructed and original signals is shown as the gray curve. (D-F) Reconstruction of a frequency-domain sparse signal with $K$ = 830, formed by a Gaussian pulse train (pulse width 120 ps, repetition rate 10 MHz, bandwidth 8.30 GHz). (D) Original input signal $x$. (E) Reconstructed signal $\hat{x}$. The horizontal axis spans 10 MHz to 8.30 GHz with 10 MHz step, corresponding to 830 nonzero sparse coefficients. (F) Reconstructed Gaussian pulse signal $\hat{x}$. The residual error between the reconstructed and original signals is shown as the gray curve. (G) System frequency response. With timing jitter of 45.3 ps, 20.2 ps, and 3.0 ps, the 3 dB bandwidths are 5.5 GHz, 14.8 GHz, and 106.2 GHz, respectively. Solid lines are simulations; colored markers are experimental data. (D) Comparison of reconstructed picosecond pulse signals with oscilloscope sampling. Pulse widths are 1000 ps, 500 ps, and 100 ps. Solid lines show oscilloscope traces; dashed lines show reconstructed signals. Oscilloscope bandwidth is 20 GHz, sampling rate 80 GSa/s. Photon count rate is ~ 1 Mcps, with a reconstruction integration time 1 s.

Although this approach does not push the sampling efficiency to the theoretical limit of adaptive sampling, it fully exploits the incoherent sampling advantage conferred by the intrinsic randomness of the quantum probe state and attains a reconstruction bandwidth far exceeding that of conventional methods. With detectors exhibiting 45.3 ps and 20.2 ps time jitter, the measured 3 dB bandwidths reach 5.5 GHz and 14.8 GHz, respectively (Fig. 2G). If advanced superconducting detector with a timing jitter of 3.0 ps is employed (*33*), simulations indicate that the bandwidth would extend to 106.2 GHz. We further validated this capability by reconstructing picosecond-scale pulse signals. Figure 2(H) compares the reconstructed waveform with that captured by a wideband oscilloscope (20 GHz bandwidth, 80 GSa/s). While the oscilloscope requires 80 GSa/s uniform sampling, our system reconstructs the pulse using only ~1 Mcps of photon count rate. A detailed analysis of the DFT scheme can be found in Supplementary Material S3.2, and a comparison with the TLS scheme is presented in Supplementary Material S3.3.

***Time-domain sparse signal measurement***

For time-domain sparse signals, the sparse domain and the measurement domain naturally coincide, eliminating the need for any additional domain-alignment operation. Steps 1 and 2, quantum probe preparation and linear signal mapping, are identical to those in the frequency-domain case, and the linearly mapped photons are directly detected in the time domain. By examining the reconstruction success rate and error under varying signal dimensions and sparsity levels, we directly verified that QCS compresses the measurement number from the classical CS bound of $M = O(K\log(N/K))$ to $M \sim K$.

***Adaptive sampling mechanism*;** To quantitatively assess the sampling efficiency advantage of QCS over classical CS, we used a Dirac pulse sequence with sparsity $K$ as the test signal, $x(t) = \Sigma_k\, a_k \cdot \delta(t - \tau_k)$, where $a_k > 0$ is the pulse amplitude and $\tau_k$ is the pulse position. In classical CS, the standard approach employs an $M \times N$ random measurement matrix $\Phi$ to obtain $y = \Phi x$, incurring a theoretical lower bound of $M = O(K\log(N/K))$. In QCS, by contrast, the measurement process is inherently signal-dependent, as each photon detection event constitutes a direct sample of a nonzero pulses, yielding an adaptive mechanism.

In the experiment, the signal period $T$ is divided into $N$ time bins. With domain alignment automatically satisfied, we perform $M$ time-resolved photon detections directly on the linearly mapped state $|\psi_x\rangle = |\alpha_x\rangle$, yielding a sequence of photon arrival times $y = \{t_1, t_2, ..., t_M\}$. The detection probability in the $n$-th time bin is proportional to the signal amplitude at that moment. Each detection event thus corresponds to a one-hot row vector with a 1 at the measured time bin, and these $M$ vectors collectively form an equivalent measurement matrix

$$\Phi = \begin{bmatrix} \phi_1 \\ \phi_2 \\ \vdots \\ \phi_M \end{bmatrix} = \begin{bmatrix} 0 & 1 & 0 & & 0 & 0 \\ 0 & \cdots & 0 & \cdots & 1 & 0 \\ & & & \vdots & & \\ 0 & 0 & 1 & & 0 & 0 \end{bmatrix}, \tag{7}$$

the position of the "1" is determined by the timestamp of the $m$-th measurement. Unlike classical CS, where $\Phi$ is fixed in advance and signal-independent (*34*), the equivalent measurement matrix $\Phi$ in QCS is inferred post hoc from the detection outcomes. Each detection event directly votes for a specific time bin, and measurement resources are automatically steered toward the nonzero time bins. This signal dependence is the physical manifestation of adaptive sampling.

***Verification of the adaptive sampling limit***; To characterize the sampling efficiency of QCS, we examined how the reconstruction success rate, the required measurement number $M$, and the reconstruction error vary with the sparsity $K$ and signal dimension $N$. Figure 3(A) shows the success rate for $N=2^{15}$ and $2^{20}$, with $K = 10$, 20, 50, and 100. When $M$ exceeds approximately $2K$, the success rate rapidly approaches unity, at which point the accumulated photon counts at the nonzero time bins become statistically distinguishable from the background and the signal support set can be reliably identified.

To understand this threshold, we examined the scaling of $M$ with $K$. As shown in Figure 3(B), the measurement number $M$ grows linearly with $K$ and remain decoupled from the signal dimension $N$. Across the examined range, they lie consistently below the classical CS theoretical lower bound $M = O(K \log(N/K))$. The reconstruction error, quantified by the normalized mean squared error, decreases monotonically with increasing $M$ and exhibits a stable convergence trend, as shown in Fig. 3(C), indicating that reconstruction accuracy can be continuously improved by accumulating more measurements (*35, 36*). Together, these results directly confirm that QCS compresses the measurement number from the classical logarithmic scaling to a constant-order scaling of $O(K)$. A detailed statistical analysis is provided in Supplementary Material S4.

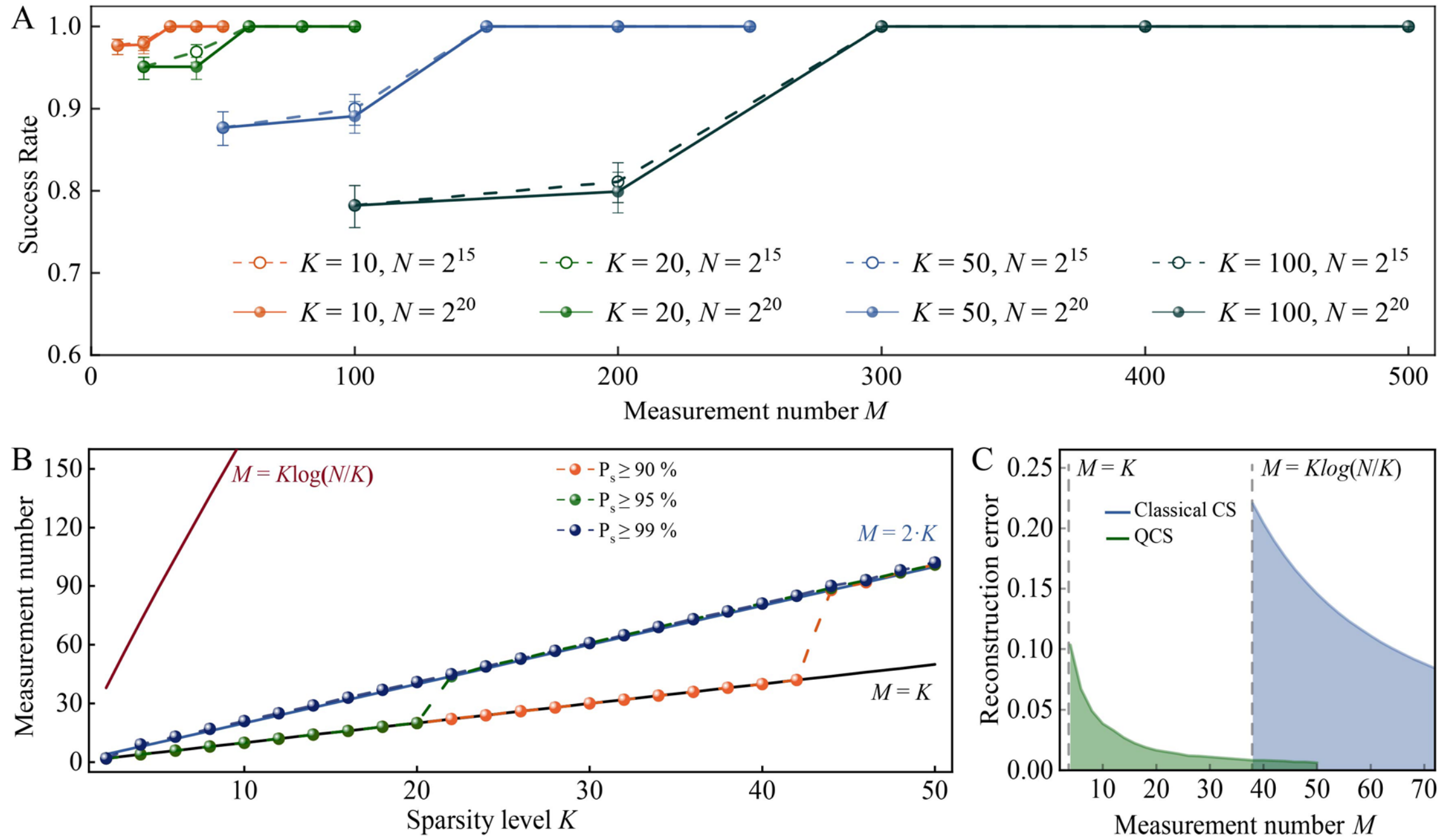

**Figure 3. QCS results for time-domain sparse signals.** (A) Reconstruction success rate versus measurement number $M$ for various signal dimensions $N$ and sparsity levels $K$. The success rate rapidly approaches unity when $M$ exceeds approximately $2K$. Error bars represent 95% confidence intervals. (B) Linear relationship between the measurement number $M$ and sparsity level $K$ in QCS. The theoretical lower bound of classical CS, $M = K\log(N/K)$, is shown for reference. The ideal limit $M = K$ and the experimentally observed scaling $M = 2K$ are also plotted as benchmarks. The signal dimension is fixed at $N = 2^{20}$. (C) Reconstruction error of CS and QCS versus measurement number $M$. The green solid line represents the experimental results of QCS, while the blue solid line shows the theoretical reference trend of classical CS. The reconstruction error exhibits a decreasing trend with increasing $M$ and approaches a stable low-error regime. The shaded regions illustrate the distinct error convergence regimes of the two approaches. The signal dimension is fixed at $N = 2^{20}$.

The frequency-domain and time-domain experiments together validate the central claim of the QCS paradigm. The former demonstrates the feasibility of transforming the mathematical reconstruction problem into a physical evolution process via domain alignment, while the latter quantitatively benchmarks the resulting leap in sampling efficiency. This leap originates from the parallel encoding capacity of quantum superposition and the collapse-based adaptive mechanism guided by domain alignment, which together constitute the fundamental advantage of QCS over classical methods.

It is worth noting that the QCS paradigm is a general theoretical framework not limited to frequency or time domain sparse signals. Scenarios such as spatial-domain imaging, high-dimensional quantum state tomography, and magnetic or electric field sensing can also be accommodated within this framework (*37-39*). Beyond coherent states, other quantum resources such as entangled states may serve as probe states, further extending the applicability of the paradigm.

**Conclusion**

We have proposed and experimentally demonstrated a quantum compressed sensing paradigm that compresses the required measurement number from the classical lower bound of $M = O(K \log(N/K))$ to $M \sim K$. This is achieved by converting the combinatorial search for the sparse support set from a computational problem into a physical evolution, using domain alignment to map the sparse basis onto the measurement basis so that projective measurements concentrate directly on the nonzero components. Experiments on frequency-domain and time-domain sparse signals validate both configurations. By shifting the computational burden from digital algorithms to quantum evolution, this approach opens a physical pathway toward the ultimate efficiency of information acquisition.

**Acknowledgements**

This work was supported by Quantum Science and Technology-National Science and Technology Major Project 2021ZD0300705; Shanxi Province Basic Research Program 202503021211084; National Natural Science Foundation of China U23A20380, 12404057, 62575162, 62127817, U25D8006 and U22A2091; Overseas Expertise Introduction Project for Discipline Innovation 111 project D18001.

**Author Contributions**

The research concept and theory were developed by J.-Y. H.; Experimental measurements and subsequent data analysis were carried out by W. L., S.-X. W. and J.-Y. H.; The initial manuscript was prepared by J.-Y. H. and W. L.; J.-Z. T. made contributions to the theory. L.-W Z. contributed to the theory and the preparation of the manuscript. All authors contributed to the critical revision and final approval of the manuscript.

**Competing interests:** Authors declare that they have no competing interests.

**Data and materials availability:** All data are available in the manuscript or in the supplementary materials.

**Supplementary Materials**

S1 Measurement number limit of classical compressed sensing

S2 Theoretical supplement to QCS

S3 Detailed experimental setup

S4 Statistical analysis of QCS time-domain sparse signal reconstruction

# Supplementary Materials for

## Quantum compressed sensing

Jianyong Hu[1,4,7]*†, Wei Li[1,4]†, Shuxiao Wu[1,4]*, Liwen Zhang[2], Yongchuang Sun[1,4], Jiazhao Tian[3], Guosheng Feng[5], Zhixing Qiao[5], Jianqiang Liu[6], Changgang Yang[1,4], Ruiyun Chen[1,4], Chengbing Qin[1,4], Guofeng Zhang[1,4], Liantuan Xiao[1,3,4,7]* and Suotang Jia[1,4]

[1]*State Key Laboratory of Quantum Optics Technologies and Devices, Institute of Laser Spectroscopy, Shanxi University, Taiyuan 030006, China*

[2]*School of Physics and Information Engineering, Shanxi Normal University, Taiyuan 030031, China*

[3]*College of Physics and Optoelectronics Engineering, Taiyuan University of Technology, Taiyuan 030024, China*

[4]*Collaborative Innovation Center of Extreme Optics, Shanxi University, Taiyuan 030006, China*

[5]*College of Medical Imaging, Shanxi Medical University, Taiyuan, 030001, China*

[6]*College of Information Engineering, Shanxi Vocational University of Engineering Science and Technology, Jinzhong 030619, China*

[7]*Hefei National Laboratory, Hefei 230088, China*

**Corresponding author E-mail address: jyhu@sxu.edu.cn; wushuxiao1@sxu.edu.cn; xlt@sxu.edu.cn*

†*These authors contributed equally to this work.*

**The PDF file includes:**

**S1 Measurement number limit of classical compressed sensing**

To establish a clear baseline for evaluating the sampling efficiency of quantum compressed sensing (QCS), we examine the fundamental lower bound on the measurement number in classical compressed sensing (CS). We first illustrate a random-sampling-based CS strategy for frequency-sparse signals and then present the general information-theoretic lower bound (*1*).

Assume that the signal $x(t)$ is $K$-sparse in the frequency domain, meaning that its discretized representation $x\in\mathbb{R}^N$ (where $N$ is determined by the Nyquist sampling rate) satisfies $x = \Psi\boldsymbol{S}$, where $\Psi$ is the sparse basis matrix (e.g., the Fourier basis) and $\boldsymbol{S}$ contains only $K$ nonzero elements. Instead of sampling uniformly at the Nyquist rate, a sampler waits for a random time sequence $\{t_i\}$ ($i$=1…$M$} to trigger each sample, yielding a measurement vector $y = \{y_1, y_2, \ldots, y_M\}$, where $M$ is the measurement number and $M \ll N$.

At the $i$-th random instant $t_i$, the acquired measurement $y_i$ can be expressed as:

$$y_i = x(t_i) = \int_{-\infty}^{\infty} x(\tau)\cdot\delta(\tau - t_i)d\tau\ . \tag{S1}$$

The sampling operation is equivalent to taking the inner product of the signal with an impulse vector $\phi_i$, which contains a 1 only at the index corresponding to $t_i$ and zeros elsewhere. All $M$ such impulse vectors together constitute the measurement matrix:

$$\Phi = \begin{bmatrix} \phi_1^T \\ \vdots \\ \phi_M^T \end{bmatrix}. \tag{S2}$$

We thus obtain the standard CS model $y = \Phi x = \Phi\Psi S$, where $\Theta = \Phi\Psi$ is the sensing matrix.

Next, we demonstrate that $\Phi$, and consequently $\Theta$, satisfies the restricted isometry property (RIP) with high probability. When the sparse basis $\Psi$ is the Fourier basis, the sensing matrix $\Theta$ is a partial Fourier matrix. (Note: In the main text, the signal model is $x = \Psi S$, with $\Psi$ being the inverse Fourier matrix. Consequently, $\Theta = \Phi\Psi$ is a partial inverse Fourier matrix. Since the Fourier matrix and the inverse are conjugate transposes of each other, their RIP properties are entirely equivalent. We refer to $\Theta$ as a partial Fourier matrix hereafter, following the standard terminology in the CS literature.) It has been rigorously proven that when the measurement number satisfies $M \geq CK\log^4(N)$, a uniformly random partial Fourier matrix satisfies the RIP with overwhelming probability (*2*).

A tighter bound for the required measurement number is $M \geq CK\log(N/K)$, which represents the currently accepted optimal bound.

**Theorem:** If an $M \times N$ matrix $\Phi$ satisfies the $K$-th order RIP (with $\delta_K$ sufficiently small), then the measurement number $M$ must satisfy $M \geq CK\log(N/K)$, where $C$ is a constant depending on $\delta_K$. This lower bound $M \geq CK\log(N/K)$ constitutes the measurement limit of classical CS (*12*).

This limit can be understood from an information-theoretic perspective. To uniquely determine a $K$-sparse signal, the total information acquired through $M$ measurements must be at least the amount of information required to describe the signal. The number of ways to choose $K$ positions out of $N$ possible locations is $C(N, K) \approx (N/K)^K$, and describing this combinatorial number requires at least $\log_2(C(N, K)) \approx K \log_2(N/K)$ bits. Consequently, $M$ must be proportional to $K\log(N/K)$, where the logarithmic factor $\log(N/K)$ represents precisely the combinatorial search cost of locating the $K$ nonzero entries among $N$ positions. This is the price that must be paid for the non-adaptive nature of classical CS.

**Why is $M = K$ impossible in classical CS?** Several fundamental constraints preclude this.

***The locality problem*;** Our goal is to recover $K$ nonzero values from $M$ measurements $y$. One knows only that a $K$-sparse signal has $K$ nonzero entries, but not where among the $N$-dimensional space they reside. If $M = K$, one has only $K$ measurement opportunities, meaning the sensing matrix $\Theta$ has only $K$ rows. To uniquely determine both the positions and the values of the $K$ nonzero coefficients, every column of $\Theta$ would need to be distinct and form a nonsingular $K \times K$ linear system. Yet the signal's support set is precisely the unknown information that CS seeks to recover.

***The RIP constraint;*** Classical CS requires the sensing matrix to satisfy the RIP. it can be shown that any matrix satisfying the $K$-th order RIP must have a measurement number substantially larger than $K$. No properly designed matrix can satisfy the RIP with $M = K$. Theoretically, $M = K$ is possible only when the support set is known *a* priori, in which case the problem reduces to a simple linear algebra task, and the very purpose of CS vanishes.

The innovation of QCS lies in its use of quantum physical properties to create a signal-dependent adaptive measurement paradigm that successfully circumvents the $O(\log(N/K))$ combinatorial search cost, thereby reducing the required measurement number to $M \sim K$. This is the theoretical origin of the quantum advantage.

**S2 Theoretical supplement to QCS**

The main text presents the four-step framework of the QCS paradigm and outlines the core function and physical implementation of each step. To maintain conciseness, certain technical details and extended discussions were omitted from the main text. This supplementary section supplements the main text in three areas. Section S2.1 presents two representative examples of linear signal-to-quantum-state mapping, phase encoding based on NV centers and intensity encoding based on single-photon imaging, to illustrate how this step can be realized in different physical systems. Section S2.2 compares the computational complexity of signal reconstruction in classical CS and QCS, and provides a systematic comparison along dimensions such as hardware resources, power consumption, and noise robustness. Section S2.3 supplements the theoretical details of using coherent states as quantum probe states in frequency-domain sparse signal measurements, including a proof that their intrinsic randomness satisfies the RIP and a derivation of the linear mapping realized by an electro-optic modulator.

**S2.1 Additional examples of linear signal-to-quantum-state mapping**

**Example 1: Phase encoding (for measuring physical fields, e.g., magnetic field sensing)**

Signal: a magnetic field strength $B$.

Quantum probe state: A qubit (e.g., the electron spin of a nitrogen-vacancy center in diamond) (*3, 4*).

Linear signal-to-quantum-state mapping process:

1. Prepare the qubit in a superposition of the ground and excited states, $|\psi_0\rangle = (|0\rangle+|1\rangle)/\sqrt{2}$, serving as the quantum probe state.

2. Allow the qubit to evolve under the influence of the magnetic field signal $B$ for a duration $\tau$. The field induces an energy-level splitting, creating a relative phase difference $\phi$ between $|0\rangle$ and $|1\rangle$.

3. The phase difference is proportional to the magnetic field strength $\phi=\gamma B\tau$, where $\gamma$ is the gyromagnetic ratio, a constant.

Encoded quantum state:

$$|\psi_x\rangle = \left(|0\rangle + e^{i\gamma B\tau}|1\rangle\right)/\sqrt{2}\,, \tag{S3}$$

Note that the expectation value $\langle\sigma_x\rangle=\cos(\phi)$ measured in the $X$ basis is nonlinear in $B$. However, through orthogonal measurements or parameter estimation techniques, the phase $\phi$ can be directly extracted from the measurement statistics, and the linear relationship $\phi = \gamma B\tau$ constitutes the physical basis for the linear mapping in QCS. Within the QCS paradigm, this linear relationship corresponds to Step 2 of the main text. Subsequently, domain-alignment evolution maps the phase information onto the measurement basis, completing the compressed sampling of the sparse signal.

**Example 2: Intensity encoding (for image classification tasks)**

Signal: A two-dimensional intensity image $I$, where each pixel value $I(u, v)$ represents its reflectance or transmittance.

Quantum probe state: A spatially coherent photon.

Linear signal-to-quantum-state mapping: The photon impinges on the image, which acts as a mask. At each pixel ($u$, $v$), the photon is transmitted (or reflected) with probability $T(u, v) \propto I(u, v)$ (*5*).

Encoded quantum state:

$$\left|\psi_x\right\rangle = \sum_{u,v} \sqrt{T\left(u,v\right)} \cdot \left|u,v\right\rangle, \tag{S4}$$

where $|u, v\rangle$ denotes the eigenstate of a photon located at position ($u$, $v$). The photon detection probability at a given position is:

$$P\left(u,v\right) = \left|\left\langle u,v \middle| \psi_x \right\rangle\right|^2 = T\left(u,v\right) \propto I\left(u,v\right), \tag{S5}$$

which is proportional to the image pixel value $I(u, v)$, corresponding to Step 2 of the main text. The image information is encoded into the spatial probability distribution of the photon. Subsequent domain-alignment evolution can be implemented using optical elements such as spatial light modulators to complete the compressed reconstruction of sparse coefficients.

**S2.2 Comparison of computational complexity for signal reconstruction in classical CS and QCS**

As discussed in the main text, domain-alignment evolution transforms signal reconstruction from an ill-posed combinatorial optimization problem into a linear parameter estimation task, reducing the reconstruction computational complexity to the $O(K)$ level. This section provides a detailed comparison between classical CS and QCS in terms of computational complexity and hardware resources, revealing the paradigm shift enabled by QCS. QCS transfers the most computationally intensive part of the reconstruction process from the digital computer to the physical system via domain-alignment, reducing the computational complexity by several orders of magnitude.

**Computational complexity of signal reconstruction in classical CS**

The core of classical CS reconstruction lies in recovering the sparsest solution (i.e., the $\ell_0$-norm minimization problem) from an underdetermined linear system $y = \Phi x$. Since $\ell_0$-norm minimization is NP-hard, it is typically relaxed to the following convex optimization problem, namely $\ell_1$-norm minimization:

$$\min_x \left\|S\right\|_1 \qquad subject \quad to \quad y = \Phi\Psi S, \tag{S6}$$

which demands substantial computational resources and specialized convex optimization or greedy iterative algorithms.

Convex optimization algorithms, such as Basis Pursuit, equivalently transform $\ell_1$-norm minimization into a linear programming problem. The complexity of solving a linear program generally scales polynomially with the problem size (number of variables $N$ and number of constraints $M$). For interior-point methods, the complexity can theoretically reach $O((N+M)^{3.5})$, which becomes computationally prohibitive for high-dimensional signals (*6*).

Greedy iterative algorithms, such as Orthogonal Matching Pursuit, iteratively select the columns (atoms) of the measurement matrix that are most correlated with the residual to progressively approximate the signal.

Each iteration requires computing the inner product of the residual with all atoms (complexity $O(MN)$), and solving a least-squares problem to update the coefficients. When $k$ atoms selected, the least-squares step costs roughly $O(k^2M)$. Completing $K$ iterations thus yields a total complexity of approximately $O(KMN+K^3)$, where the $K^3$ term dominates for high sparsity $K$ (28). Classical reconstruction is therefore an inherently computation-intensive, iterative search executed entirely in the digital domain, with its burden growing polynomially with the signal dimension $N$ and sparsity $K$.

**Computational complexity of signal reconstruction in QCS**

Signal reconstruction in QCS exhibits markedly different computational characteristics. When the measurement domain and the sparse domain are perfectly aligned, the reconstruction process is non-iterative. The sparse coefficient amplitudes can be obtained directly from statistics of the measurement statistics $\hat{s}_n = k_n/(\eta M)$, where $k_n$ is the count in the $n$-th measurement basis state, $M$ is the measurement number, and $\eta$ is a global scaling factor. QCS reconstruction merely requires counting the $M$ measurement outcomes across the $N$ measurement basis states, an operation of computational complexity $O(M)$. Since $M \sim K$, the reconstruction complexity is equivalently $O(K)$. This linear-time process represents one of the least computationally expensive paradigms in signal processing. Table S1 provides a systematic comparison.

**Table S1. Comparison of classical CS and QCS**

| Feature | Classical CS | QCS |
|---|---|---|
| Nature of Reconstruction | Combinatorial search and sparse-constrained optimization | Direct statistical estimation |
| Core Algorithm | Linear Programming, Greedy Pursuit | Arithmetic averaging |
| Measurement Number $M$ | $M = O(K\log(N/K))$ | $M \sim K$ |
| Reconstruction Complexity | $O((N+M)^{3.5})$ or $O(KMN+K^3)$ | $O(K)$ |
| Computational Bottleneck | Signal dimension $N$ and sparsity $K$ | None |
| Computational Resources | Hardware: High-performance CPU/GPU clusters<br>Memory: ~20 GB (storing $\Phi$ matrix, $N = 2^{20}$)<br>Power: ~200-500 W (typical GPU server)<br>Time: ~10-100 ms (software execution) | Hardware: Single-photon detector + TDC<br>Memory: ~1 kB (storing counters)<br>Power: ~10 W (core detection unit)<br>Time: ns-μs scale (domain-alignment evolution; total measurement time proportional to $M$) |
| Where Computation Occurs | Digital computer (software algorithm) | Physical system (domain alignment) + digital computer (lightweight post-processing) |
| Scalability | Memory-limited; difficult for $N > 10^5$ on standard computers | Memory independent of $N$ |
| Noise Robustness | Reconstruction is an ill-posed inverse problem, sensitive to noise; stable recovery requires $M \geq O(K\log(N/K))$ to ensure RIP. | "Voting" mechanism converts reconstruction into a linear estimation problem; noise manifests as a uniform background, suppressible statistically by increasing $M$ with $\sim 1/\sqrt{M}$ scaling. |

QCS transfers the computational burden traditionally borne by software/algorithms onto the physical hardware layer, directly substituting complex computations with quantum physical evolution. This paradigm shift yields quantum advantages in three key aspects. First, a fundamental leap in measurement efficiency: the required measurement number is reduced from the classical lower bound of $M = O(K\log(N/K))$ to $M \sim O(K)$, approaching the adaptive sampling limit—a transition from logarithmic to constant-order dependence. Second, an exponential reduction in computational complexity: back-end processing is simplified from exponential iterative optimization in the classical case to constant-order statistical counting. Third, inherent advantages in noise resilience: QCS transforms an ill-posed nonlinear inverse problem into a well-posed linear parameter estimation problem. The effect of noise can be suppressed by simply increasing the measurement number according to the $1/\sqrt{M}$ statistical scaling, whereas classical CS reconstruction algorithms exhibit intrinsic instability in the presence of noise.

**S2.3 Supplement to frequency-domain sparse signal measurement based on QCS**

This supplementary section provides the quantum optical theoretical foundation for the frequency-domain sparse signal measurement experiment described in the main text. It includes the statistical description of coherent states, a proof that their intrinsic randomness satisfies the RIP, and a quantum optical derivation of the linear mapping realized by an electro-optic modulator.

**S2.3.1 Quantum probe state—Description of coherent states**

The experiment in the main text employs a coherent state $|\alpha\rangle$ as the quantum probe state. This subsection supplements its quantum optical description, providing the theoretical basis for the intrinsic randomness that satisfies the RIP.

A coherent state $|\alpha\rangle$ is an eigenstate of the annihilation operator $\hat{a}$, satisfying $\hat{a}|\alpha\rangle = \alpha|\alpha\rangle$, where $\alpha=|\alpha|e^{(i\theta)}$ and the mean photon number is $\langle n\rangle=\langle\alpha|\hat{a}^{\dagger}\hat{a}\,|\alpha\rangle=|\alpha|^2$. Its expansion in the Fock basis is:

$$|\alpha\rangle = e^{-|\alpha|^2/2}\sum_{n=0}^{\infty}\frac{\alpha^n}{\sqrt{n!}}|n\rangle, \tag{S7}$$

where $|n\rangle$ is the Fock state with $n$ photons. The photon counting distribution obeys Poisson statistics:

$$P(n) = |\langle n \mid \alpha\rangle|^2 = e^{-\langle n\rangle}\frac{\langle n\rangle^n}{n!}, \tag{S8}$$

Optical attenuation can be modeled by a beam splitter with transmittance $\eta$ ($0 \le \eta \le 1$). A coherent state $|\alpha\rangle$ and a vacuum state $|0\rangle$ are incident on the beam splitter; the output transmitted mode remains a coherent state $|\sqrt{\eta}\alpha\rangle$ with reduced amplitude. The attenuated weak coherent state then impinges on an ideal photon-number-resolving detector. According to the Born rule, the probability of detecting $k$ photons is

$$P(k) = \left|\left\langle k \middle| \sqrt{\eta}\alpha\right\rangle\right|^2 = e^{-\eta|\alpha|^2}\frac{\left(\eta|\alpha|^2\right)^k}{k!}. \tag{S9}$$

In the continuous-spectrum case (e.g., position operator), the probability density is $|\psi(x)|^2$, reflecting the same fundamental randomness. For a photon-number-non-resolving single-photon detector (click/no-click detector), the measurement operators are $\Pi_0=|0\rangle\langle 0|$ and $\Pi_{\text{click}}=I-|0\rangle\langle 0|$. The click probability is $P(\text{click})=1-exp(-\eta|\alpha|^2)$. Although the detector cannot resolve exact photon numbers, the temporal randomness of click events originates from the same Poisson statistics: the click events constitute a Poisson point process, with exponentially distributed inter-arrival times and no memory. This intrinsic randomness is the physical foundation for the universal sensing capability of QCS.

**S2.3.2 Guarantee of universal sensing in QCS**

Building on the Poisson statistics and click probability derived in S2.3.1, we now prove that this intrinsic randomness guarantees the RIP—the mathematical condition that ensures QCS can serve as a universal sensing framework for arbitrary input signals.

In the main text, we noted that to ensure universal sensing, the eigenbasis of the quantum probe state must match the original domain of the signal (e.g., time), and the probe state must exhibit a uniform probability distribution in that basis. The randomness of photon arrival times from a coherent state naturally provides such a uniform distribution. The core task of this subsection is to build a bridge connecting the physical phenomenon of "randomness in photon arrival times" to the mathematical theorem that "random matrices satisfy the RIP."

Divide the total measurement time $T$ into $N$ time bins of duration $\delta t$ ($T = N\delta t$). The time bin $\delta t$ must be sufficiently short so that within each bin, the probability of detecting more than one photon is much smaller than the probability of detecting exactly one photon (i.e., the single-photon condition). In this regime, the Poisson process within a single time bin approximates a Bernoulli trial, with the detection probability in the $n$-th bin given by $p_n = \lambda_n\delta t \ll 1$, where $\lambda_n$ is the mean photon count rate. The detection outcome is thus a random variable $y_n$ taking the value 1 (detection) or 0 (no detection).

The measurement process in QCS can be represented by a matrix $\Phi\in\{0,1\}^{M\times N}$, where each row corresponds to a single photon detection event and contains exactly one 1 at the time bin where the photon arrived, with zeros elsewhere. For any signal $x\in\mathbb{R}^N$, the $m$-th measurement yields $(\Phi x)_m = x_{n(m)}$, where $n(m)$ is the position of the 1 in the $m$-th row. This is equivalent to sampling the components of x$x$ with replacement, where each row independently selects a position with uniform probability.

Let $Y = \|\Phi x\|_2^2 = \sum_{m=1}^{M} x_{n(m)}^2$ represents the average squared norm of the measured signal components. Because each row independently selects a position with uniform probability $1/N$, its expectation is:

$$E[Y] = \sum_{m=1}^{M} E\left[x_{n(m)}^2\right] = M\cdot\left(\frac{1}{N}\sum_{n=1}^{N} x_n^2\right) = \frac{M}{N}\|x\|_2^2. \tag{S10}$$

The corresponding variance is:

$$Var(Y) = \sum_{m=1}^{M} Var\left(x_{n(m)}^2\right) = M\cdot\left(\frac{1}{N}\sum_{n=1}^{N} x_n^4 - \left(\frac{1}{N}\sum_{n=1}^{N} x_n^2\right)^2\right). \tag{S11}$$

For a $K$-sparse vector $x$, there are at most $K$ nonzero components. Let the nonzero components be $x_{n1}$, ..., $x_{nK}$, then:

$$\|x\|_2^2=\sum_{k=1}^{K}x_{j_k}^2\ ,\ \ \sum_{j=1}^{N}x_j^4=\sum_{k=1}^{K}x_{j_k}^4\ . \tag{S12}$$

By the Cauchy-Schwarz inequality $\sum_{k=1}^{K}x_{j_k}^4\le\left(\sum_{k=1}^{K}x_{j_k}^2\right)^2=\|x\|_2^4$, hence:

$$Var(Y)\le M\cdot\left(\frac{1}{N}\|x\|_2^4-\frac{1}{N^2}\|x\|_2^4\right)=\frac{M(N-1)}{N^2}\|x\|_2^4\ . \tag{S13}$$

Next, we need to show that $Y$ is close to its expectation $\frac{M}{N}\|x\|_2^2$. Since $Y$ is a sum of independent random variables, with $x_{j(i)}^2\in\left[0,\|x\|_\infty^2\right]$ and $\|x\|_\infty^2\le\|x\|_2^2$. By Bernstein's inequality, we have:

$$P\left(\left|Y-\frac{M}{N}\|x\|_2^2\right|\ge t\right)\le 2\exp\left(-\frac{t^2/2}{Var(Y)+bt/3}\right), \tag{S14}$$

where $b=\|x\|_\infty^2\le\|x\|_2^2$. Let $t=\delta\frac{M}{N}\|x\|_2^2$, then:

$$P\left(\left|Y-\frac{M}{N}\|x\|_2^2\right|\ge\delta\frac{M}{N}\|x\|_2^2\right)\le 2\exp\left(-\frac{\left(\delta M\|x\|_2^2/N\right)^2/2}{\frac{M(N-1)}{N^2}\|x\|_2^4+\frac{\delta M\|x\|_2^2/N\cdot\|x\|_2^2}{3}}\right). \tag{S15}$$

Simplify the denominator of the exponent:

$$\frac{M(N-1)}{N^2}\|x\|_2^4+\frac{\delta M}{3N}\|x\|_2^4=M\|x\|_2^4\left(\frac{(N-1)}{N^2}+\frac{\delta}{3N}\right). \tag{S16}$$

Thus:

$$P\left(\left|S-\frac{M}{N}\|x\|_2^2\right|\ge\delta\frac{M}{N}\|x\|_2^2\right)\le 2\exp\left(-\frac{\delta^2M^2\|x\|_2^4/\left(2N^2\right)}{M\|x\|_2^4\left(\frac{N-1}{N^2}+\frac{\delta}{3N}\right)}\right)=2\exp\left(-\frac{\delta^2M/2}{(N-1)/N+\delta/3}\right).$$

(S17)

For large $N$ we have $(N-1)/N\approx 1$, so:

$$P\left(\left|\|\Phi x\|_2^2-\frac{M}{N}\|x\|_2^2\right|\ge\delta\frac{M}{N}\|x\|_2^2\right)\le 2\exp\left(-c\delta^2M\right), \tag{S18}$$

where $c$ is a constant.

The above probability bound holds for a fixed $x$. We require it to hold uniformly for all $K$-sparse vectors. Consider an $\varepsilon$-net: there exists an $\varepsilon$-net covering the unit sphere of $K$-sparse vectors, with size at most $\binom{N}{K}(3/\epsilon)^K$ . Taking $\varepsilon$ = 0.1, the net size is $\le \binom{N}{K}30^K$ . The union bound requires $2\binom{N}{K}30^K \exp\left(-c\delta^2 M\right) < \eta$ , i.e., $M \ge \frac{1}{c\delta^2}\left(K\log(N/K) + K\log(30) + \log(1/\eta)\right)$ . Therefore, as long as $M \ge CK\log(N/K)$ (with $C$ depending on $\delta$ and $\eta$), we have with high probability:

$$\left| \|\Phi x\|_2^2 - \frac{M}{N}\|x\|_2^2 \right| \le \delta \frac{M}{N}\|x\|_2^2, \tag{S19}$$

that is:

$$(1-\delta)\frac{M}{N}\|x\|_2^2 \le \|\Phi x\|_2^2 \le (1+\delta)\frac{M}{N}\|x\|_2^2. \tag{S20}$$

Letting $\delta K = \delta$, the normalized matrix $\frac{\sqrt{N}}{\sqrt{M}}\Phi$ satisfies the RIP:

$$(1-\delta_K)\|x\|_2^2 \le \left\| \frac{\sqrt{N}}{\sqrt{M}}\Phi x \right\|_2^2 \le (1+\delta_K)\|x\|_2^2, \tag{S21}$$

The original matrix $\Phi$ satisfies the RIP with constant $\delta K$ (though its expectation is $M/N$, requiring normalization).

We have thus proved that when $M \ge O(K\log(N/K))$, the normalized random sampling matrix $\frac{\sqrt{N}}{\sqrt{M}}\Phi$ satisfies the RIP with high probability (hence $\Phi$ itself satisfies the RIP, up to a scaling constant). The RIP for the original matrix $\Phi$ takes the form:

$$(1-\delta_K)\frac{M}{N}\|x\|_2^2 \le \|\Phi x\|_2^2 \le (1+\delta_K)\frac{M}{N}\|x\|_2^2. \tag{S22}$$

This proof establishes that the coherent-state–based random sampling matrix inherits the RIP from the intrinsic Poisson randomness of photon arrivals, thereby guaranteeing the universal sensing capability of QCS for arbitrary input signals (*11*).

**S2.3.3 Linear signal-to-quantum-state mapping**

In the frequency-domain measurement experiment of the main text, a broadband electro-optic intensity modulator linearly maps the radio-frequency signal onto the optical field. This subsection provides the quantum optical derivation, showing that the photon detection probability of the output optical field is linearly related to the input radio-frequency signal.

Consider a Mach–Zehnder (MZ) intensity modulator consisting of a 50:50 beam splitter, two phase-modulation arms, and a beam combiner. The input is a coherent state described by the annihilation operator $\hat{a}$. After the first beam splitter, the field operators for the two arms are $\hat{a}_1 = \hat{a}_2 = \hat{a}\sqrt{2}$. Applying a time-varying phase $\varphi_1(t)$ to one arm and a fixed phase $\varphi_2=0$ to the other, the modulated operators become $\hat{a}_1'=\exp(i\varphi_1(t))\,\hat{a}_1$ and $\hat{a}_2' = \hat{a}_2$. The output mode of interest (one combiner port) is

$$\hat{a}_{out} = \frac{1}{\sqrt{2}}\left(\hat{a}_1^{'} + \hat{a}_2^{'}\right) = \frac{e^{i\varphi_1(t)} + 1}{2}\hat{a} \tag{S23}$$

The relative phase difference is $\Delta\varphi(t)=\varphi_1(t)$. The output state remains a coherent state $|\psi_x\rangle$ with eigenvalue $\alpha x(t)$, where the modulation function $x(t)=(e^{i\Delta\varphi(t)}+1)/2$. The photon detection probability at the output is proportional to $|\alpha x(t)|^2$, which is linearly related to the intensity modulation of the input radio-frequency signal when the modulator is biased at the quadrature point and operated in the small-signal regime. This directly corresponds to the linear signal-to-quantum-state mapping (Step 2 of QCS).

**S2.4 Quantum information-theoretic explanation of QCS**

The main text states that QCS compresses the measurement number from the classical CS bound of $M = O(K\log(N/K))$ to $M \sim K$. This section provides a rigorous theoretical explanation for this leap from the perspective of quantum information theory and clarifies the physical boundaries of the QCS advantage.

**S2.4.1 The binding of information acquisition and measurement in classical information theory**

The lower bound on the measurement number in classical CS, $M = O(K\log(N/K))$, stems from a fundamental information-theoretic fact. Fully describing a $K$-sparse signal requires two types of information. First, positional information: the locations of the $K$ nonzero components in an $N$-dimensional space, whose uncertainty requires approximately $K\log(N/K)$ bits. Second, amplitude information: the numerical values of the $K$ nonzero coefficients, requiring roughly $Kb$ bits (where $b$ is the quantization bit depth per coefficient). The non-adaptive measurement mechanism of classical CS requires the measurement matrix to treat all possible positions uniformly; the information extracted from each measurement must simultaneously cover both positional and amplitude dimensions. Since the information extractable from a single measurement is limited by the channel capacity, acquiring a total of $K\log(N/K) + Kb$ bits inevitably demands a measurement number proportional to the total information content, leading to the bound $M = O(K \log(N/K))$. The logarithmic term $\log(N/K)$ is precisely the measurement cost that must be paid to acquire positional information.

This analysis rests on the implicit assumption that information can only be obtained through measurement data. In classical physics, this is natural—there is no unitary evolution, and any information processing must be extracted via measurement. However, quantum systems permit unitary evolution prior to measurement, and unitary evolution itself can reorganize and transform information without consuming measurement trials. Classical information theory cannot describe this mechanism of "processing information via evolution," which constitutes a fundamental limitation of its framework.

**S2.4.2 Relationship between the adaptive sampling limit and the information-theoretic limit**

Two concepts are often conflated but must be distinguished. The **information-theoretic limit** (Shannon's source coding theorem) is the minimum information required to describe a $K$-sparse signal, approximately $K\log(N/K) + Kb$ bits. This is an absolute lower bound from the perspective of source coding—regardless of the measurement or coding strategy employed, this much information must be acquired. The **adaptive sampling limit** is the measurement number needed when the support set is known a priori: only $K$ targeted measurements are required to acquire the amplitudes of the $K$ non-zero coefficients. This limit on the measurement number does not violate the information-theoretic limit, because the positional information was supplied as prior knowledge rather than extracted through measurement. The reason the adaptive sampling limit lies below the classical CS measurement lower bound is that positional information acquisition is shifted from the "measurement phase" to the "prior knowledge phase."

**S2.4.3 The information payment structure of QCS: unitary evolution as an information-processing resource**

The core innovation of QCS lies in its use of domain-alignment evolution—a physically realizable unitary transformation—to identify the support set at the quantum level without converting it into classical bits or consuming measurement trials. Concretely, if the signal is sparse in the basis $\Psi$, the domain-alignment evolution $\hat{U}_e$ satisfies

$$\widehat{U}_e \left\{ \left| \psi_n \right\rangle \right\} = \left\{ \left| \omega_n \right\rangle \right\}, \tag{S24}$$

mapping the sparse basis $\{|\psi_n\rangle\}$ onto the measurement basis $\{|\omega_n\rangle\}$. This unitary transformation "folds" the signal's sparse positional information into the probability amplitude structure of the quantum state. Subsequent projective measurements merely sample from this pre-aligned distribution to acquire the amplitude information of each nonzero component.

From an information-theoretic perspective, QCS alters the payment structure of information. In classical CS, both positional and amplitude information must be paid for by the measurement data, resulting in a measurement lower bound of $M = O(K\log(N/K))$. In QCS, positional information is absorbed by the domain-alignment evolution, consuming no measurement trials; only amplitude information must be acquired through measurement, reducing the measurement lower bound to $M \sim K$. The total information remains conserved, but the means of payment has been fundamentally altered.

Note that this offset is possible because we know the signal's sparse domain in advance (e.g., frequency domain). This prior knowledge is also exploited in classical CS—one must know $\Psi$ to construct the sensing matrix—but classical CS still pays the $\log(N/K)$ measurement cost due to the non-adaptive nature of $\Phi$. QCS physically embeds this prior knowledge into the measurement process via unitary evolution, thereby exempting the measurement number from this cost.

**S2.4.4 The Holevo bound and the measurement number lower bound of QCS**

The Holevo bound constrains the amount of classical information extractable from a quantum state via measurement. For a quantum system in a state $\rho = \Sigma_x p_x \rho_x$, the mutual information between the measurement outcome $Y$ and preparation index $X$ satisfies

$$I\left(X;Y\right) \le S\left(\rho\right) - \sum_x p_x S\left(\rho_x\right), \tag{S25}$$

where $S(\cdot)$ is the von Neumann entropy. This implies that the classical information extractable from a single measurement is finite, upper-bounded by the entropy of the quantum state.

In QCS, the output state $|\psi_{out}\rangle$ after domain-alignment evolution is an $N$-dimensional quantum state. Even if it encodes all the signal information, a single projective measurement yields only one classical outcome, from which at most $\log_2 N$ bits can be extracted. To fully acquire the amplitude information of the $K$ nonzero components, each requires accumulating sufficient photon counts to suppress statistical fluctuations. Experiments show that, on average, approximately two effective detection counts per nonzero component are needed for stable identification (see Supplementary Material S4 for details). Consequently, the lower bound on the measurement number is $O(K)$, consistent with the Holevo bound. This limit cannot be further reduced by optimizing the unitary evolution; it is a physical boundary set by the fundamental principles of quantum mechanics.

**S2.4.5 Perspective from quantum channel capacity**

Domain-alignment evolution can be formalized as a quantum channel, with the input being the quantum state encoding the signal and the output being a quantum state amenable to measurement. Its classical capacity is given by the Holevo information

$$C(\mathcal{E}) = \max_{\{p_x,\rho_x\}}\left[S\left(\mathcal{E}\left(\sum_x p_x\rho_x\right)\right) - \sum_x p_x S\left(\mathcal{E}\left(\rho_x\right)\right)\right]. \tag{S26}$$

However, the efficiency advantage of QCS does not stem from increasing the classical capacity, but from the fact that the channel itself performs a computational task—basis transformation from the sparse basis to the measurement basis—that would require $O(N\log N)$ operations on a classical computer. The associated combinatorial search for the support set of $K$ non-zero components among $N$ positions has NP-hard complexity. A quantum system accomplishes this transformation in a single unitary evolution, independent of $N$, a quintessential manifestation of quantum parallelism in signal processing.

**S2.4.6 Quantum fisher information and parameter estimation precision**

From the perspective of quantum parameter estimation, QCS aims to estimate the $K$ nonzero parameters of the sparse coefficient vector $S$. The covariance matrix of the estimates is constrained by the quantum Cramér-Rao bound

$$Cov\left(\hat{S}\right) \geq \frac{1}{M} F_Q^{-1}, \tag{S27}$$

where $F_Q$ is the quantum Fisher information matrix. Crucially, domain-alignment evolution concentrates the quantum Fisher information onto the measurement basis states corresponding to the $K$ nonzero components, making the rank of $F_Q$ approximately $K$ with its non-zero eigenvalues corresponding to the individual sparse components. Therefore, only $M \sim K$ measurements are needed to saturate the quantum Cramér-Rao bound and achieve optimal parameter estimation precision. Classical CS, unable to concentrate the Fisher information in advance, requires $O(K\log(N/K))$ measurements for comparable estimation accuracy.

**S2.4.7 Physical boundaries and limits of QCS**

Based on the above analysis, the physical boundaries of the QCS advantage can be clearly delineated.

First, the upper bound on absorbable positional information is at most $K\log(N/K)$ bits, corresponding to the combinatorial uncertainty of the sparse support set. This offset presupposes prior knowledge of the signal's sparse domain; If the sparse domain is unknown, QCS reverts to an ordinary measurement, and the measurement number returns to $O(N)$.

Second, amplitude information cannot be absorbed. It is the content carried by the signal itself, varying from one measurement instance to another, and must be acquired through projective measurements. The Holevo bound and the quantum Fisher information dictate that acquiring amplitude information requires at least $O(K)$ measurements.

Third, the information extraction from a single measurement is bounded. Even after domain-alignment evolution, a single projective measurement yields at most $\log_2 N$ bits. This is a direct consequence of the postulates of quantum measurement and cannot be circumvented by different unitary evolutions. To fully reconstruct the $K$ non-zero components, a sufficient measurement number must be accumulated.

**S2.4.8 Summary**

QCS achieves sparse signal reconstruction with $M \sim K$ measurements. Its essence lies in altering the payment structure of information acquisition. In classical CS, both positional and amplitude information must be paid for by the measurement data, yielding a measurement lower bound of $O(K\log(N/K))$. In QCS, positional information is absorbed at the quantum level by domain-alignment evolution, and the measurement data need only pay for amplitude information, reducing the measurement lower bound to $O(K)$. This alteration of the payment structure is possible because the unitary evolution of a quantum system is itself an information-processing resource—it can reorganize and transform the encoded information without consuming measurement trials.

The physical essence of QCS can be distilled as: parallel exploration through quantum superposition and adaptive focusing through projective collapse. A single quantum probe state simultaneously traverses all possible sparse configurations in the signal space before measurement. Domain-alignment evolution physically maps the results of this parallel exploration onto the measurement basis. When projective measurement occurs, wavefunction collapse spontaneously steers toward the non-zero components with higher probability. Each collapse event is itself an adaptive sample performed without prior knowledge. The combinatorial search cost is thus dissolved by quantum parallelism and the adaptive mechanism of collapse. This perspective situates QCS within the unified theoretical framework of quantum sampling, quantum machine learning, and other quantum-advantage protocols.

## S3 Detailed experimental setup

### S3.1 Experimental setup for frequency-domain sparse signal QCS

This section details the experimental setup for frequency-domain sparse signal QCS. The core is a quantum optical system that physically implements domain alignment, mapping the frequency-domain sparse information onto time-domain photon detection events. The setup comprises four functional modules corresponding to the four steps of the QCS paradigm, as shown in Figure S1. The key component parameters are listed in Table S2.

A continuous-wave laser (LTSS, DFB-1550-10-PM-FA-M) with a central wavelength of 1550 nm generated a coherent state $|\alpha\rangle$ as the quantum probe state (QCS Step 1). A $LiNbO_3$ intensity modulator (bandwidth 30 GHz) linearly mapped the radio-frequency signal $x(t)$ under test onto the optical field (QCS Step 2). The DC bias was set to the quadrature point to ensure operation in the linear regime, yielding a modulated intensity $I(t) \propto (1 + m{\cdot}x(t))$, where $m$ is the modulation depth, thus establishing a linear relationship between the signal and the photon detection probability.

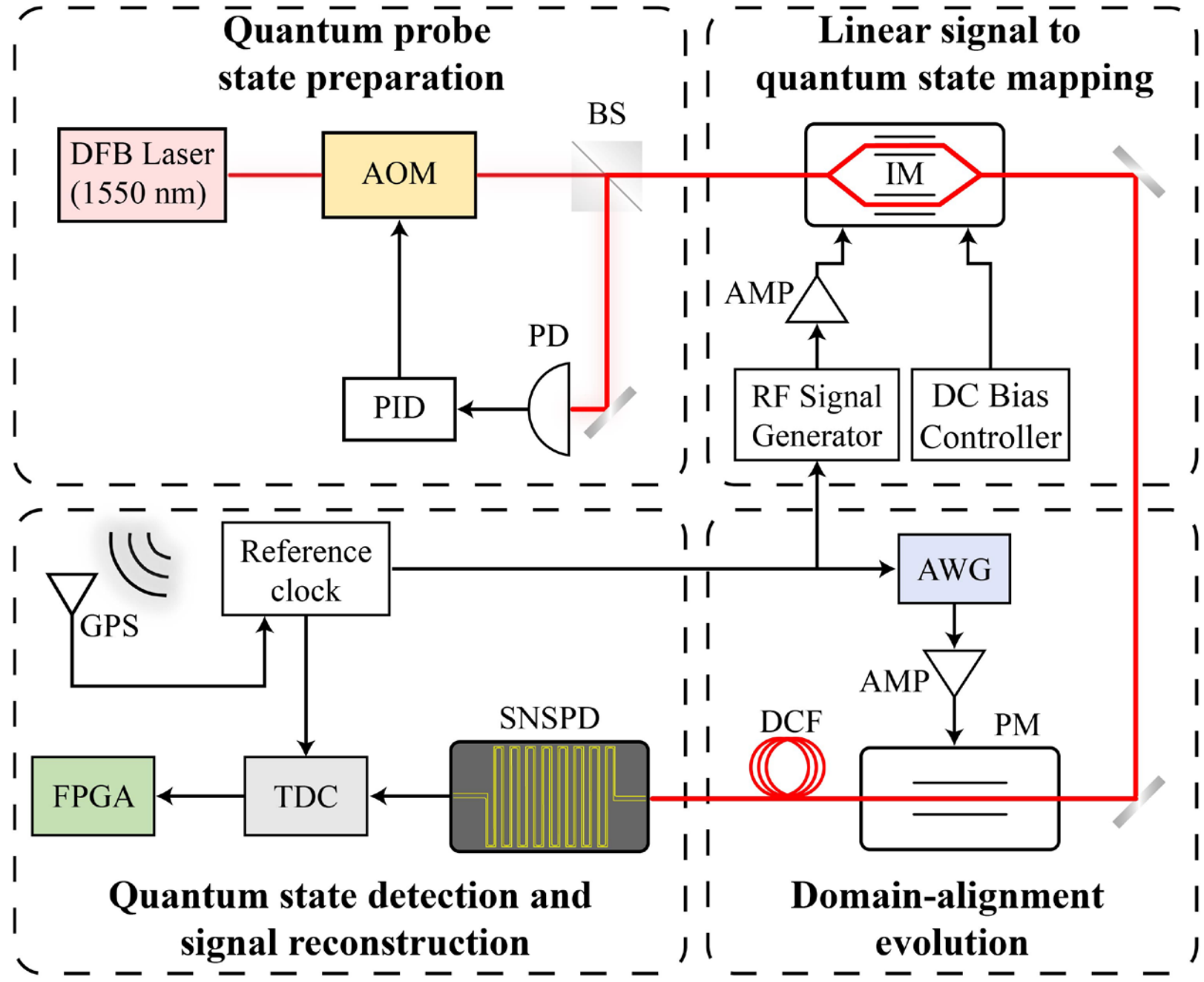

**Figure S1. Experimental setup for frequency-domain sparse signal QCS measurement.** The four functional modules are labeled with their corresponding QCS steps. Step 1: Laser generates the coherent state probe. Step 2: Intensity modulator linearly maps the RF signal onto the optical field. Step 3: Time lens performs domain-alignment evolution. Step 4: Single-photon detector and TDC perform detection and timestamp recording, followed by post-processing for signal reconstruction. AOM: acousto-optic Modulator; BS: beam splitter; PD: photoelectric detector; PID: Proportional-integral-differential controller; IM: intensity modulator; AMP: amplifier; PM: phase modulator; AWG: arbitrary waveform generator; DCF: dispersion compensation optical fiber; SNSPD: superconducting single photon detector; TDC: time-to-digital converter; FPGA: field programmable gate array; GPS: Global Positioning System.

Domain-alignment evolution (QCS Step 3) was implemented by a time-lens spectrometer (TLS), consisting of an electro-optic phase modulator cascaded with a dispersion-compensating fiber. The phase modulator (bandwidth 25 GHz) was driven by a periodic parabolic radio-frequency signal, imposing a quadratic temporal phase $\varphi(t) \propto t^2$ within each period $T$, which introduced a linear chirp. The dispersion-compensating fiber provided a second-order dispersion of $\ddot{\phi} = 1074\ \text{ps}^2$, satisfying the imaging condition $C_L\ddot{\phi}=1$ together with the chirp rate of the time lens, where $C_L$ is the lens strength. This combination physically realizes a time-domain Fourier transform within each time window. Consequently, the probability $P(t_n)$ of detecting a photon at time $t_n$ is proportional to the power spectral density $|S(f_n)|^2$ of the input signal at frequency $f_n$, with the linear time-to-frequency mapping $f_n=-t_n/(2\pi\ddot{\phi})$. Thus, the frequency-domain sparse coefficients of the signal are aligned onto the detection time basis.

Photon detection employed a superconducting nanowire single-photon detector (QCS Step 4) with a peak system detection efficiency of 93% at 1550 nm and a dark count rate below 10 cps. Photon arrival timestamps were recorded in real time by a time-to-digital converter (TDC) (time resolution 1 ps). The timestamps were binned into corresponding frequency bins, directly yielding estimates of the sparse coefficient amplitudes $\hat{s}_n$. Full waveform reconstruction is then completed via the inverse transform $\hat{x} = \Psi\hat{\boldsymbol{S}}$.

**Table S2. Key experimental component parameters**

| Module | Component | Key Parameters | Function and Remarks |
|---|---|---|---|
| **Light Source** | CW laser (Model: DFB-1550-10-PM-FA-M) | Center wavelength: 1550 nm<br>Output power: 10 mW<br>Linewidth: 3 MHz | Provides the coherent quantum probe state. Narrow linewidth ensures temporal coherence, essential for high-precision interference and dispersion processing. |
| **Modulation and Encoding** | Intensity modulator (Model: AX-6K5-30-PFA-PFAP-R6) | Bandwidth: 30 GHz<br>Half-wave voltage ($V_\pi$): 4 V<br>Insertion loss: 5 dB<br>Extinction ratio: > 60 dB | Linearly maps the RF signal under test onto the optical field. High bandwidth and low $V_\pi$ facilitate high-frequency, high-linearity modulation. |
| **Time Lens (Domain Alignment)** | Electro-optic phase modulator (Model: PM-DK5-40-PFA-PFA-LV) | Bandwidth: 25 GHz<br>$V_\pi$ (RF): 4 V | Applies quadratic phase modulation to realize the "lens" effect for time-to-frequency conversion. Requires synchronization with the driving RF signal. |
| **Dispersion Processing (Domain Alignment)** | Dispersion-compensating fiber (Model: ADCM-SH-C-100%-050-FC/APC-03B) | Dispersion (DL): -843 ps/nm<br>Second-order dispersion ($\ddot{\phi} = \beta_2 L$): 1074 ps$^2$<br>Length ($L$): 5.2 km | Implements the dispersive part of the time-lens spectrometer, linearly mapping frequency-domain information onto the time domain. Dispersion must match the modulation signal bandwidth. |

| **Single-Photon Detection** | Superconducting nanowire single-photon detector (Model: Single Quantum Eos 800 CS) | Maximum efficiency: 93% @ 1550 nm<br>Dark count rate: < 10 cps | Converts weak photon events into electrical pulses. |
|---|---|---|---|
| **Time-to-Digital Conversion** | Time-to-Digital Converter (Model: Time Tagger Ultra) | Time resolution: 1 ps<br>Maximum count rate: 70 Mcps | Accurately records photon arrival timestamps. High resolution (ps-level) is crucial for system bandwidth and precision. |
| **Synchronization and Clock** | GPS-disciplined clock (Model: SYN3307) | Short-term stability: $5\times10^{-11}$ @ 1 s<br>Long-term stability: ≤ $1\times10^{-12}$ (24 h average)<br>Phase noise: ≤ -110 dBc/Hz @ 10 Hz, ≤ -145 dBc/Hz @ 10 kHz | Provides a low-jitter, high-stability synchronization reference clock for the signal, time-lens driver, and TDC. |

**S3.2 System bandwidth limitation and frequency resolution**

The primary factor limiting the system's reconstruction bandwidth is the timing jitter of the single-photon detector. This section first models the timing jitter, then quantifies its impact on measurement bandwidth and frequency resolution, and finally demonstrates multi-component reconstruction capability under these constraints.

**Timing jitter model.** The system timing jitter $\Delta t$ arises from the convolution of the timing jitters of the SPAD and the TDC. It typically follows an exponentially modified Gaussian distribution:

$$P(t) = \frac{h}{\sqrt{2\pi}\sigma} erfcx\left(\frac{\sigma}{\sqrt{2}\tau} + \frac{t\text{-}\mu}{\sqrt{2}\sigma}\right)\cdot \exp\left(-\frac{(t\text{-}\mu)^2}{2\sigma^2}\right), \tag{S28}$$

where $\mu$ is the mean time delay, $\sigma$ is the standard deviation of the Gaussian component, $\tau$ is the time constant of the exponential decay, $h$ is a normalization factor, and $erfcx(x) = e^{x2}\cdot erfc(x)$ is the scaled complementary error function. This distribution indicates that the system jitter is composed of a Gaussian kernel $\exp(-(t-\mu)^2/(2\sigma^2))$ and an asymmetric exponential correction $erfcx(\cdot)$.

In the experiment, two types of detectors were employed, giving system timing jitters of 45.3 ps and 20.2 ps, respectively. Figure S2(A) shows the measured jitter histograms together with fits to the exponentially modified Gaussian function; the green dashed line shows a simulation for a superconducting nanowire detector with 3.0 ps jitter.

**Bandwidth limitation.** In broadband radio-frequency signal measurement, the effect of timing jitter is equivalent to that of a low-pass filter. Taking the Fourier transform of the timing jitter distribution yields the system's frequency response. The amplitude response $|H(f)|$ is dominated by the Gaussian term and modulated

by the exponential tail:

$$\left|H(f)\right| \approx \exp\left(-2\pi^2\sigma^2 f^2\right)\cdot\left[1+\left(2\pi\tau f\right)^2\right]^{-1/2}. \tag{S29}$$

The effective measurement bandwidth of the system is usually defined as the -3 dB bandwidth $f_{3\text{dB}}$. Setting $|H(f_{3\text{dB}})|=1/\sqrt{2}$ gives an approximate inverse proportionality to the full width at half maximum of the jitter distribution $\Delta t_{\text{FWHM}}$:

$$f_{3dB} \approx \frac{C}{\Delta t_{FWHM}}, \tag{S30}$$

where the constant $C$ depends on the shape of the distribution (the ratio of $\sigma$ to $\tau$). For a nearly Gaussian distribution ($\tau \rightarrow 0$), $C \approx 0.3$. In the experiment, the exact bandwidth values were obtained by direct Fourier transform of the measured jitter distributions. With the 45.3 ps and 20.2 ps detectors, the corresponding -3 dB bandwidths are 5.5 GHz and 14.8 GHz, respectively; simulations indicate that a 3 ps jitter would extend the bandwidth to 106.2 GHz. These results are consistent with Figure 2G of the main text and confirm that detector timing jitter is the core physical factor limiting the reconstruction bandwidth.

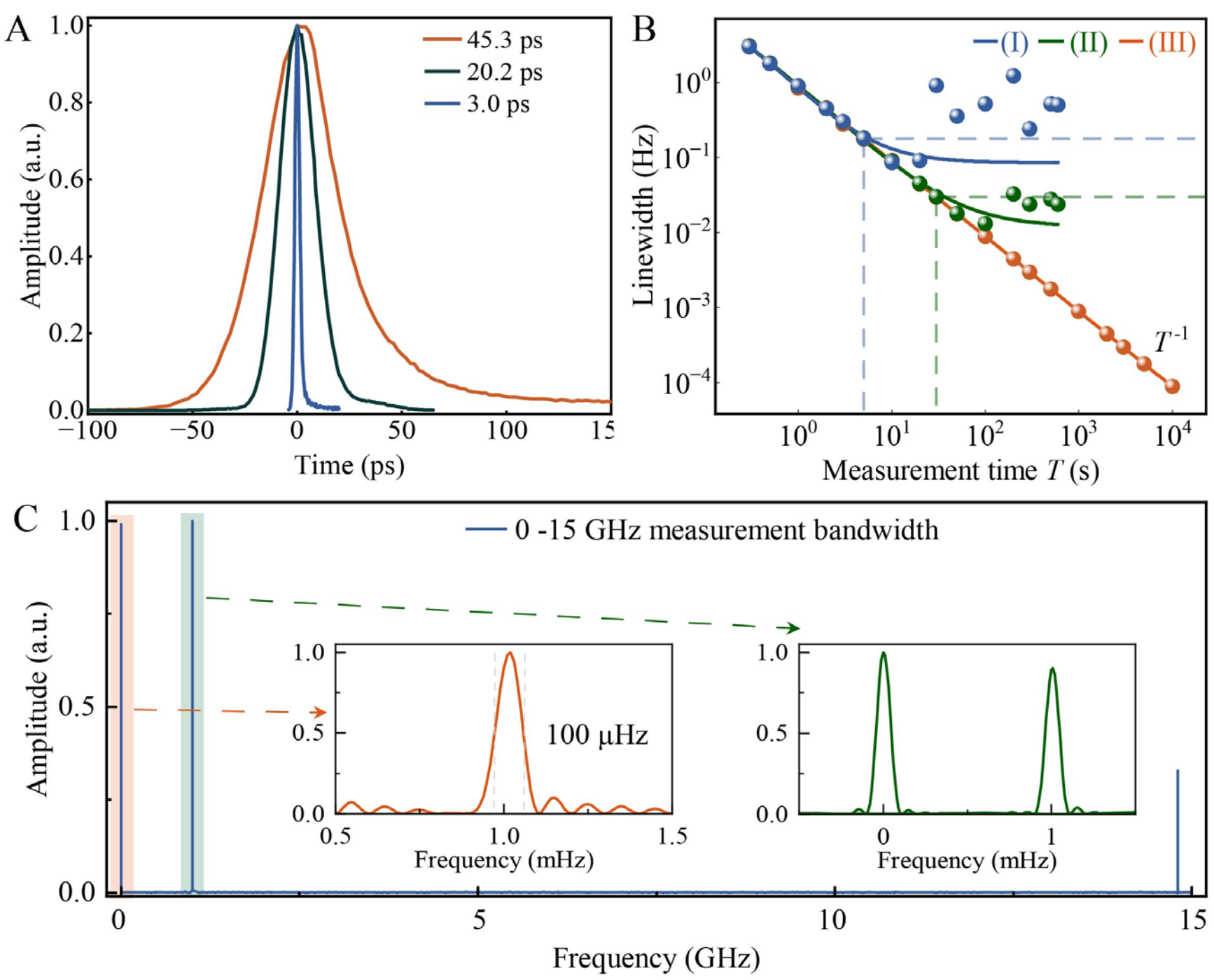

**Figure S2. Performance for frequency-domain sparse signal QCS with DFT scheme.** (A) Characterization of system timing jitter. Blue and orange line represent experimental measured timing jitter distributions for the two detectors (system jitter 45.3 ps and 20.2 ps, respectively); The blue line curve corresponds to the simulated result for a superconducting detector with 3 ps timing jitter. (B) Frequency resolution characterization. Curve I: independent clocks for the signal source and TDC; Curve II: both clocks locked to a GPS standard; Curve III: ideal Fourier-limited condition with a common clock. The test signal is a 1.0 GHz single-tone sine wave. (C) Reconstruction of a multi-component sparse signal containing four frequency components (1.0 mHz, 1.0 GHz, 1.0 GHz + 1.0 mHz, and 14.8 GHz). Integration time: 10,000 s, yielding a frequency resolution of 100 μHz. Insets: magnified views of the 1.0 mHz component and the closely spaced 1.0 GHz / (1.0 GHz + 1.0 mHz) components.

**Frequency resolution.** While timing jitter limits the absolute bandwidth, frequency resolution determines the ability to distinguish adjacent sparse components. Figure S2B shows the frequency resolution as a function of integration time for three clock configurations. Curve III corresponds to the ideal Fourier-limited condition, where the signal source and the TDC share a common clock; the frequency resolution equals the reciprocal of the integration time $T^{-1}$. Curve I corresponds to independent clocks for the signal source and TDC; the resolution is then limited by the relative frequency drift, with a measured linewidth of approximately 5.0 Hz. Curve II corresponds to both clocks locked to a GPS standard; the residual frequency error after locking is 0.03 Hz. These results demonstrate that, with appropriate clock synchronization, ultra-high frequency resolution can be achieved independently of the bandwidth limitation imposed by timing jitter.

**Multi-component reconstruction.** The ability to simultaneously reconstruct multiple components directly verifies the system's capacity for handling K>1$K$>1 sparse signals. Figure S2C shows the reconstruction of four frequency components (1.0 mHz, 1.0 GHz, 1.0 GHz + 1.0 mHz, and 14.8 GHz), using an integration time of 10,000 s to achieve a frequency resolution of 100 μHz. This result confirms that the system reliably resolves multi-component sparse signals under realistic bandwidth and resolution constraints.

**S3.3 The two implementation pathways for frequency-domain sparse signal measurement**

The frequency-domain sparse signal measurement presented in the main text encompasses two complementary pathways: the direct identification scheme based on time-lens spectrometer (TLS) domain alignment, and the post-processing reconstruction scheme based on the discrete Fourier transform (DFT). This section supplements the technical details of both schemes, discusses their performance differences, and clarifies their respective roles within the QCS paradigm.

***TLS domain-alignment scheme.*** The TLS implements the domain-alignment evolution operator $\hat{U}_e$ required by Equation (1) of the main text. Its core principle exploits the quadratic phase modulation imposed by an electro-optic phase modulator together with the second-order dispersion of an optical fiber to realize a Fourier transform, linearly mapping the spectral information of the input signal onto the temporal envelope of the output optical field. This process strictly follows the four steps of the QCS paradigm and serves as a physical instantiation of the domain-alignment concept.

Experimental results demonstrate that this scheme can achieve frequency identification at the single-photon level, with identification accuracy rising rapidly as the number of detected photons increases (Figure 1(C) of the main text), confirming the efficacy of adaptive sampling guided by domain alignment. However, in the current experiment, the reconstruction quality is constrained by existing engineering factors. The frequency resolution $\Delta f$ of the TLS is jointly determined by the dispersion and the time window $T$. Limited by the dispersion-compensating fiber length employed and the driving power of the commercial phase modulator, the system dispersion is 1074 $ps^2$, resulting in a frequency resolution on the order of GHz for typical time windows, which is insufficient to resolve closely spaced frequency components. Additionally, the dispersion-compensating fiber introduces higher-order dispersion and polarization mode dispersion, causing the Fourier transform to deviate from ideal conditions and producing systematic mapping errors. It should be emphasized that these limitations arise from component-level performance boundaries, not from the domain-alignment principle itself. With advances in low-jitter single-photon detectors, dispersion engineering, and integrated photonic time lenses, a clear path exists for improving this scheme.

***DFT post-processing scheme.*** The DFT scheme is identical to the TLS scheme in Step 1 (quantum probe state preparation) and Step 2 (linear signal-to-quantum-state mapping). The difference lies in omitting the physical domain-alignment evolution of Step 3; instead, a DFT is applied directly to the photon arrival time sequence to estimate the spectrum.

This scheme holds significance in two respects. First, it verifies the independent contribution of the initial QCS steps. The high-quality reconstruction results of the DFT scheme (Figs. 2A–H of the main text) demonstrate that even without physical domain alignment, the measurement data produced by Steps 1 and 2—incoherent sampling based on the intrinsic randomness of coherent-state photon arrival times—already possess an information structure far superior to that of conventional uniform sampling, indirectly confirming the independent value of the quantum probe state preparation and linear mapping steps. Second, it provides a practical pathway under current technological conditions. While the engineering implementation of physical domain alignment awaits further optimization, the DFT scheme substitutes classical post-processing for physical evolution, achieving reconstruction accuracy comparable to state-of-the-art commercial oscilloscopes (Fig. 2H of the main text) while retaining the core sampling advantages of QCS. This trade-off demonstrates the flexibility of the QCS framework.

***Modularity of the QCS paradigm.*** The modular design of QCS endows it with generality beyond any specific physical implementation. The domain-alignment evolution, as the core operator of Step 3, can be realized through different means depending on the application: when ultimate sampling efficiency is desired, physical implementations such as TLS, diffractive neural networks, or programmable quantum gate sequences can perform the unitary transformation; when engineering constraints dominate, classical spectral estimation algorithms in Step 4 can serve as an equivalent substitute. Regardless of the implementation, as long as Steps 1 and 2 guarantee incoherent sampling based on quantum intrinsic randomness, the sampling efficiency advantage of the QCS paradigm is realized to varying degrees. Physical domain alignment pushes the advantage to the theoretical limit of $M \sim K$, while classical post-processing trades a portion of the sampling efficiency for optimal reconstruction fidelity under current engineering constraints.

In summary, the TLS and the DFT scheme serve complementary roles in this work. The former confirms the theoretical attainability of pushing the measurement count to the adaptive sampling limit within the QCS paradigm, and its current experimental limitations identify directions for future engineering optimization. The latter confirms the practical efficacy of the paradigm, demonstrating that even with imperfect domain alignment, the intrinsic randomness of the quantum probe state already confers advantages unavailable to classical methods. Together, these two pathways constitute a multi-faceted validation of the QCS paradigm.

**S4 Statistical analysis of QCS time-domain sparse signal reconstruction**

This section provides a quantitative statistical analysis, based on a probabilistic model, of the relationship between the reconstruction success rate, the measurement number $M$, and the signal sparsity $K$ in the QCS scheme. It also derives the statistical scaling of the minimum required measurement number with sparsity.

Consider a signal of duration $T$, divided into $N$ time bins ($N$ sufficiently large). The signal is $K$-sparse in the time domain, meaning only $K$ time bins contain nonzero amplitudes (set to 1 for simplicity), while the remaining $N-K$ bins have zero amplitude, with $K \ll N$. This signal intensity-modulates a continuous-wave laser, and the modulated light is received by a photon-number-non-resolving single-photon detector. Under ideal conditions, each nonzero pulse induces a valid detection event with probability $p$ ($0 < p \leq 1$), and a single pulse produces at most one detection. The signal of length $T$ is played cyclically until the detector accumulates a total of $M$ photon detection events. Each recorded photon is tagged with its arrival time, thereby corresponding to a specific time bin. Reconstruction is defined as successful if all $K$ nonzero time bins are hit at least once among the $M$ recorded events.

**S4.1 High detection probability limit ($p \approx 1$)**

When the per-pulse detection probability $p = 1$, every nonzero pulse is detected. Within one signal period $T$, the detector will register $K$ photon events, each corresponding to one of the $K$ nonzero time bins. Therefore, a measurement number of $M = K$ guarantees that all $K$ nonzero time bins are hit at least once, and reconstruction is assuredly successful.

When $p$ is slightly less than 1, missed detections occur, and to cover all $K$ nonzero time bins with high probability, the required measurement number $M$ must exceed $K$. In practical QCS experiments, dark counts and background noise are also present. To enhance the statistical contrast between signal and background, a “double-count background suppression” strategy is typically employed: true nonzero time bins will generate repeated detection events, whereas random noise is unlikely to produce two or more events in the same time bin. Thus, with $p$ close to 1, a measurement number of $M$ slightly larger than $2K$ causes the reconstruction success rate to rapidly approach unity (see Figure 3(C) of the main text).

**S4.2 Statistical model under finite detection probability ($0 < p < 1$)**

When $p$ is significantly less than 1, missed detections cannot be neglected, and the reconstruction success must be described by a statistical covering process. We first construct an analytical model for $K = 2$, then generalize to $K = 3$ and arbitrary $K$.

**S4.2.1 Analytical results for the $K = 2$**

Suppose the signal contains only two nonzero time bins, A and B. Reconstruction is successful if both A and B appear at least once among the $M$ detections.

***Case of $M = 2$.*** Without loss of generality, assume the first photon falls in bin A. Subsequent detection events occur in periodic order: $B_1$, $A_2$, $B_2$, $A_3$, $B_3$, $A_4$,… Success requires that at least one of the subsequent events hits bin B:

$$P_{success}(M=2) = p + (1-p)^2 p + (1-p)^4 p + ... = \frac{1}{2-p}. \quad \text{(S31)}$$

***Generalization to arbitrary M.*** When the measurement number $M \geq 2$, failure means that after the first photon falls in some bin (e.g., A), all of the subsequent $M-1$ photons fall in the same bin A. Since successive detections are conditionally independent, the conditional probability of remaining in the same bin in a single trial is $P_{fail}(M=2)$. Hence, the reconstruction success rate is

$$P_{success}(M) = 1 - [P_{fail}(M=2)]^{M-1} = 1 - \left(\frac{1-p}{2-p}\right)^{(M-1)}. \quad \text{(S32)}$$

As $M$ increases, the success rate converges to unity exponentially. Figure S4 compares the experimentally measured reconstruction success rates (scatter points) with the simulation results (solid lines) for different values of $p$.

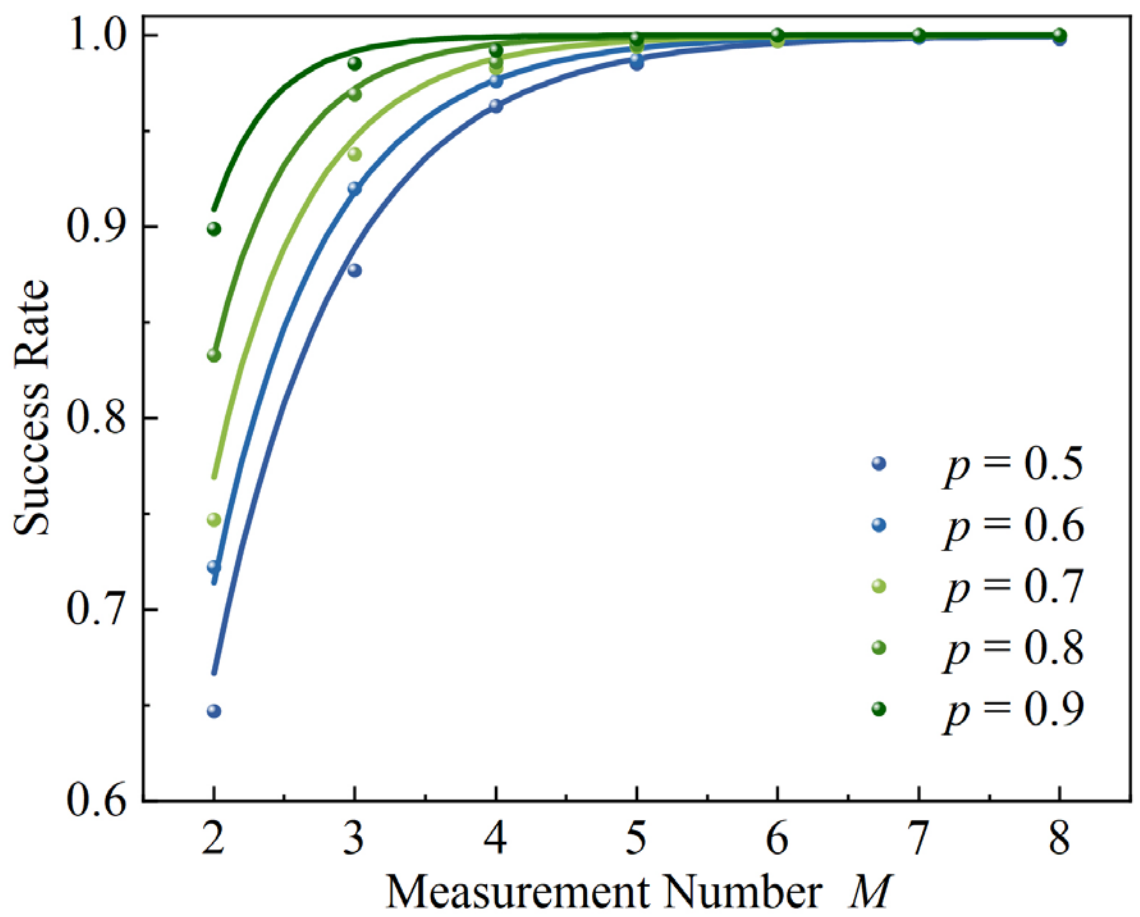

**Figure S4. Reconstruction success rate versus measurement number for different effective detection probabilities *p*.**

### S4.2.2 Statistical derivation for the $K = 3$

For $K = 3$, the signal contains three time-bins, denoted A, B, and C. Successful reconstruction requires that all three bins are detected at least once. Missed detections alter the relative distances between uncovered bins and the current position, necessitating a discrete Markov chain model.

The per-bin detection probability is $p$. The probability of at least one detection occurring in one period is $P_{period} = 1-(1-p)^3$. The conditional probability that the next detection event spans $d$ time bins ($d = 1, 2, 3$) relative to the current position is:

$$u = \frac{p}{P_{period}}, (d=1),$$

$$v = \frac{(1-p)p}{P_{period}}, (d=2),$$

$$w = \frac{(1-p)^2 p}{P_{period}}, (d=3), \quad \text{(S33)}$$

One can verify that $u+v+w=1$. Define three incomplete states, $X$: only 1 time bin has been hit so far; $Y_A$: 2 bins have been hit, and the missing bin is adjacent to the current position; $Y_B$: 2 bins have been hit, and the missing bin is two steps away. Once all three bins have been hit, the process enters the absorbing state (success). The transition matrix $T$ among the transient states (determined by the jump probabilities) is:

$$\mathrm{T}=\begin{bmatrix} w & 0 & 0 \\ u & w & u \\ v & v & w \end{bmatrix}, \tag{S34}$$

The initial state vector after the first photon is:

$$v_{init}=\begin{bmatrix} 1 \\ 0 \\ 0 \end{bmatrix}. \tag{S35}$$

If sfter accumulating a measurement number $M$, the system is still in an incomplete state, reconstruction has failed. Therefore, the failure probability is:

$$P_{fail}\left(M\right)=\begin{bmatrix}1 & 1 & 1\end{bmatrix}\mathbf{T}^{(M-1)}v_{init}. \tag{S36}$$

The final success probability is:

$$P_{success}\left(M,K=3\right)=1-\begin{bmatrix}1 & 1 & 1\end{bmatrix}\mathbf{T}^{(M-1)}v_{init}, \tag{S37}$$

As $p\rightarrow 1$, $u\rightarrow 1$, $v\rightarrow 0$, $w\rightarrow 0$, and the system rapidly covers all time bins. Figure S5 shows the experimentally measured reconstruction success rate versus measurement number for different $p$, along with simulations based on this model.

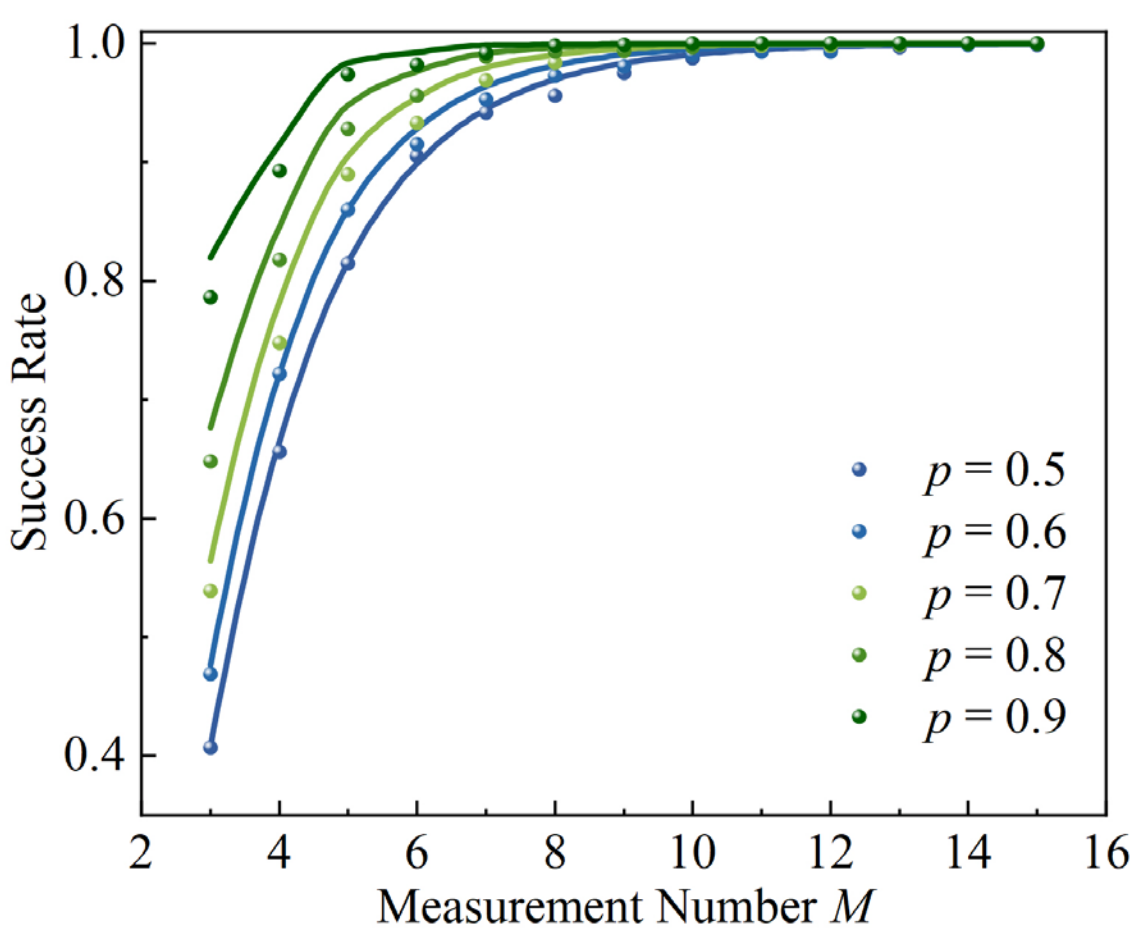

**Figure S5. Reconstruction success rate versus measurement number for different *p*.**

### S4.2.3 Generalization to arbitrary *K*

For arbitrary sparsity $K$, the probability of at least one detection occurring in a full period is $P_{period}=1\text{-}(1\text{-}p)^K$. If a photon is at a given time bin, the conditional probability that the next detection spans $d$ time bins ($d$ = 1, 2, ..., $K$, potentially crossing multiple periods) is

$$P_d = \frac{(1-p)^{(d-1)}p}{P_{period}}, \qquad \sum_{d=1}^{\infty} P_d = 1, \tag{S38}$$

The incomplete-reconstruction states are characterized by the number of covered bins $m$ ($1 \le m \le K-1$) and the relative distances between the current position and the uncovered bins. All transient states form a discrete Markov chain, with the absorbing state being the one in which all $K$ bins are covered. These form a discrete Markov chain with an absorbing state corresponding to all $K$ bins covered.

In the $K = 3$ case above, the transition probabilities were normalized for a single-period analysis; for arbitrary $K$, a similar normalization ensures that the transition matrix $T_K$ among the incomplete states is a proper stochastic matrix (rows sum to 1). Let $v_{\text{init}}$ be the initial state vector. Then

$$P_{fail}(M) = \left\| v_{init} T_K^{(M-1)} \right\|_1,$$

$$P_{success}(M) = 1 - P_{fail}(M). \tag{S39}$$

Although the state space grows with $K$, experiments indicate an approximately linear relationship between the minimum measurement number $M_{\min}$ and sparsity $K$:

$$M_{min} \approx \alpha(p)K + c \tag{S40}$$

where $\alpha(p) \ge 1$ and increases monotonically as $p$ decreases. As $p \to 1$, $\alpha(p) \to 2$, recovering the approximate relationship $M \approx 2K + c$. Figure S6 shows the experimentally measured $M_{\min}$ as a function of sparsity $K$ for different $p$, demonstrating a good linear relationship for all $p$, with slope gradually increasing as $p$ decreases. The experimental results demonstrate that the measurement system preserves a linear measurement scaling under different detection efficiency conditions.

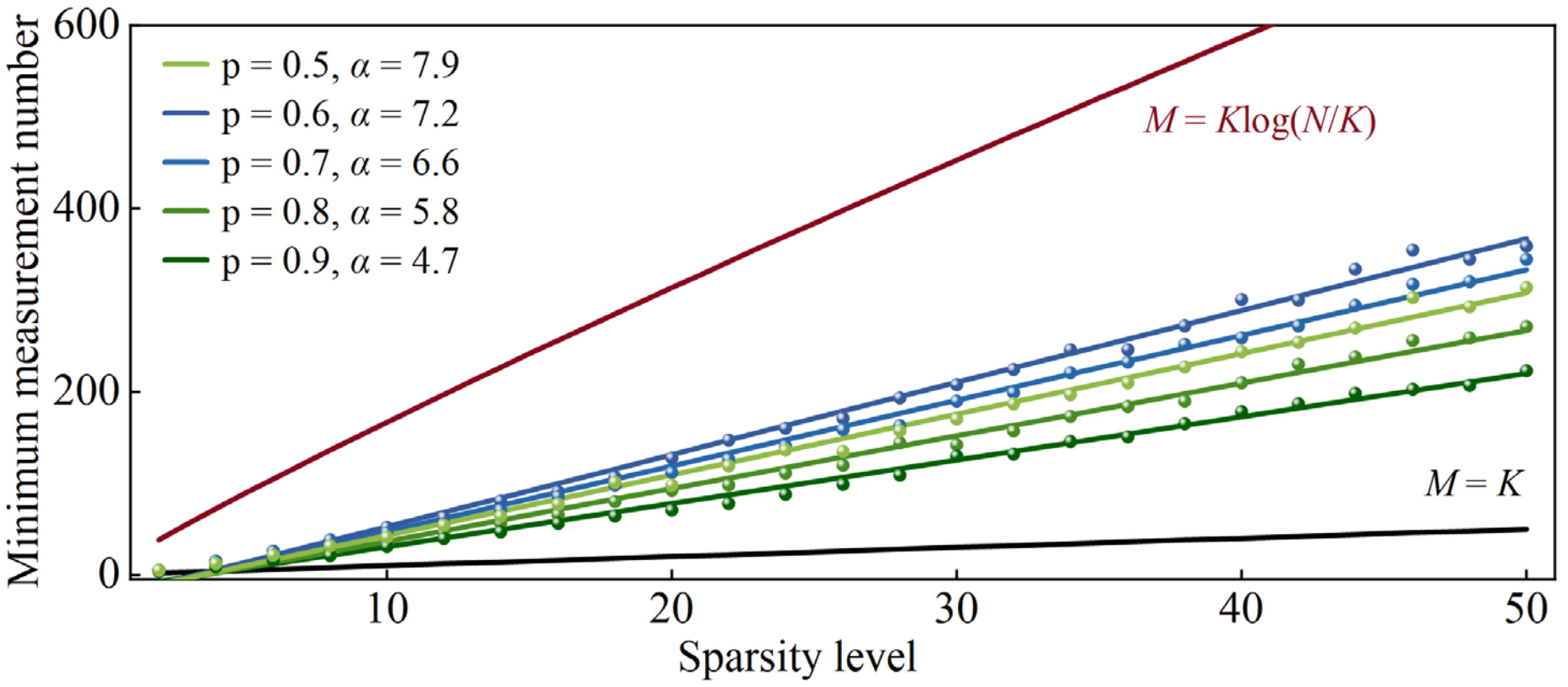

**Figure S6. Relationship between minimum measurement number and sparsity for different detection probabilities *p*.** Experimental data are shown as circular markers, with linear fits used to extract the scaling coefficients (see legend). The results exhibit an approximately linear dependence of $M$ on $K$ for all $p$. For comparison, the classical compressed sensing limit $M = K\log(N/K)$ and the ideal bound $M = K$ are also plotted.